\pdfoutput=1
\documentclass[journal,draftclsnofoot,onecolumn,12pt,twoside]{IEEEtranTCOM}
\normalsize

\ifCLASSINFOpdf
\else
\fi

\usepackage{cite}
\usepackage{amsmath,amssymb,amsfonts}
\usepackage{mathtools}
\usepackage{algorithm}
\usepackage{algpseudocode}
\usepackage{setspace}
\usepackage{graphicx}
\usepackage[caption=false,lofdepth,lotdepth]{subfig}
\usepackage{capt-of}
\usepackage{textcomp}
\usepackage{xcolor}
\usepackage{multirow}
\usepackage{gensymb}
\def\BibTeX{{\rm B\kern-.05em{\sc i\kern-.025em b}\kern-.08em
    T\kern-.1667em\lower.7ex\hbox{E}\kern-.125emX}}

\makeatletter
\newcommand\fs@betterruled{%
  \def\@fs@cfont{\bfseries}\let\@fs@capt\floatc@ruled
  \def\@fs@pre{\vspace*{5pt}\hrule height.8pt depth0pt \kern2pt}%
  \def\@fs@post{\kern2pt\hrule\relax}%
  \def\@fs@mid{\kern2pt\hrule\kern2pt}%
  \let\@fs@iftopcapt\iftrue}
\floatstyle{betterruled}
\restylefloat{algorithm}
\makeatother    

\begin{document}

\renewcommand{\algorithmicrequire}{\textbf{Input:}}
\renewcommand{\algorithmicensure}{\textbf{Output:}}
\let\oldReturn\Return
\renewcommand{\Return}{\State\oldReturn}

\renewcommand{\arraystretch}{1.25}


\title{An Efficient Code-Domain NOMA Transceiver for mm-Wave Hybrid Massive MIMO Architectures}
%
%
%

\author{Murat~Bayraktar,~\IEEEmembership{Student Member,~IEEE,}
        Gokhan~M.~Guvensen,~\IEEEmembership{Member,~IEEE}
\thanks{The authors are with the Dept. of Electrical and Electronics Engineering, Middle East Technical University, Ankara, Turkey (e-mail: muratbay@metu.edu.tr and guvensen@metu.edu.tr)}}

\maketitle
\vspace{-17mm}
\begin{abstract}
Massive MIMO and non-orthogonal multiple access (NOMA) are crucial methods for future wireless systems as they provide many advantages over conventional systems. Power-domain NOMA methods are investigated in massive MIMO systems, whereas there is little work on integration of code-domain NOMA and massive MIMO which is the subject of this study. We propose a general framework employing user-grouping based hybrid beamforming architecture for mm-wave massive MIMO systems where NOMA is considered as an intra-group process. It is shown that classical receivers of sparse code multiple access (SCMA) and multi-user shared access (MUSA) can be directly adapted. Additionally, a novel receiver architecture which is an improvement over classical one is proposed for uplink MUSA. This receiver makes MUSA preferable over SCMA for uplink transmission with lower complexity. We provide a lower bound on achievable information rate (AIR) as a performance measure. We show that code-domain NOMA schemes outperform conventional methods with very limited number of radio frequency (RF) chains where users are spatially close to each other. Furthermore, we provide an analysis in terms of bit-error rate and AIR under different code length and overloading scenarios for uplink transmission where flexible structure of MUSA is exploited.
\end{abstract}



%
\IEEEpeerreviewmaketitle

\section{Introduction}

Non-orthogonal multiple access (NOMA) is a key technology for future wireless networks due to its ability to support massive connectivity \cite{SurveyNOMA,SurveyNOMA2,SurveyNOMA3}. Unlike conventional orthogonal multiple access methods NOMA uses non-orthogonal physical resources (e.g., time and frequency) where overloading can be realized. There are two main classes of current methods: power-domain and code-domain NOMA. Sparse code multiple access (SCMA) and multi-user shared access (MUSA) are promising code-domain NOMA schemes introduced in \cite{SCMA} and \cite{MUSAforIOT}, respectively.

On the other hand, massive MIMO technology will play a crucial role in fifth generation (5G) and beyond since it offers high spectral efficiency especially at mm-wave frequencies where large frequency bands are available \cite{WhatWill_5G}. Although fully digital massive MIMO systems are proven to attain excellent performance, practical implementation of this architecture with huge number of antennas may not be possible due to increased complexity and energy requirements \cite{mMIMOoverview}. There is an alternative solution where hybrid digital-analog architecture is used which reduce the dimension of the channels in analog domain and simplify the processing by using less number of radio frequency (RF) chains \cite{Molisch1}. Joint spatial division and multiplexing (JSDM) is a two-stage beamforming based hybrid architecture that groups users according to their channel covariance eigenspaces \cite{JSDM}. Base station (BS) observes a sparse angle-delay profile at mm-wave frequencies since users have small number of multipath components (MPCs) and their angle of arrivals (AoAs) are constrained in a small angular sector which makes it possible to group users. Generalized eigen-beamformer (GEB) is a useful statistical beamforming technique based on JSDM for suppressing inter-group interference in the analog domain \cite{GMGuvensen,Anil}.

\subsection{Related Work}

NOMA schemes are initially introduced for single-antenna systems and their extension to MIMO systems is an important research direction. In addition, low-complexity interference cancellation algorithms should be developed so that NOMA schemes become more practical \cite{SurveyNOMA3}. High complexity of message passing algorithm (MPA) and error propagation of successive interference cancellation (SIC) are two main problems of current receiver architectures. There are some studies that propose novel low-complexity receivers. Conventional and advanced receivers for some NOMA schemes are described in \cite{vaezi} and references therein. In \cite{ReceiverMIMO-SCMA}, graph based iterative receivers for downlink MIMO-SCMA system with cooperation is proposed and performance comparison with existing receivers is provided where number of antennas is equal to number of physical resources. Overloaded and coded massive MIMO-NOMA systems and capacity-achieving low-complexity LMMSE detection based receivers are studied in \cite{PracticalMIMO-NOMA,CapAchMIMO-NOMA}. Gaussian message passing algorithm is proposed for overloaded massive MIMO system in \cite{GMP-MIMO-NOMA}.

In the literature, there are many studies on integration of power-domain NOMA and massive MIMO \cite{SurveyNOMA2,SurveyNOMA3}. Adaptation of NOMA to mm-wave massive MIMO systems is also studied recently. In \cite{BeamspaceNOMA}, NOMA is integrated with beamspace MIMO in order to support more users than the available RF chains. In \cite{MultiBeamNOMA}, users are divided into groups where each group has one RF chain. Multiple analog beams are created for each NOMA group by dynamically dividing antenna array into subarrays, which is called beam splitting. In \cite{mmWaveNOMA}, a joint optimization of user-grouping and hybrid beamformer design along with power allocation is provided.

On the other hand, there is limited work where code-domain NOMA is considered with massive MIMO systems. In \cite{Bjornson1} and \cite{Bjornson2}, it is shown that code-domain NOMA schemes can be useful when users are spatially close to each other in fully digital massive MIMO systems without employing a practical NOMA scheme. It is shown that grouping users according to their spatial properties increases the spectral efficiency in \cite{Bjornson2}. Massive MIMO is integrated with interleave division multiple access (IDMA) along with iterative data-aided channel estimation and suboptimal detection methods in \cite{mMIMO-IDMA}. Low density spreading (LDS) signatures are applied in massive MIMO system in \cite{mMIMO-LDS}. Belief propagation based receiver is applied without using any channel state information (CSI). In \cite{MIMO-SCMA}, beamspace MIMO at mm-wave bands is integrated with SCMA where the number of RF chains is equal to the number of physical resources.

Although there is some effort to combine NOMA with massive MIMO systems, there is lack of literature for the adaptation of practical NOMA schemes such as MUSA to massive MIMO hybrid beamforming architecture employing single-carrier in wideband mm-wave channels. In this paper, our aim is to present how NOMA can be useful in practical massive MIMO systems with limited number of RF chains where users are closely spaced.

\subsection{Contributions}

We provide a generic unification of hybrid beamforming and code-domain NOMA schemes for massive MIMO systems where the number of RF chains is limited, generally much smaller than the number of active users. To the author's knowledge, there is no such prior work that makes a complete integration of user-grouping based hybrid beamforming and NOMA schemes where NOMA is considered as intra-group separation of users which are closely located to each other. Existing code-domain NOMA schemes can be integrated with no cost of additional receiver complexity, through an equivalent system model. 

Code-domain NOMA schemes are generally considered with multicarrier transmission since inter-symbol interference (ISI) is a serious problem for single-carrier systems \cite{SurveyNOMA2,SurveyNOMA3}. In addition, most of the existing researches for massive MIMO systems adopt flat-fading channel model by considering the use of orthogonal frequency division multiplexing (OFDM) \cite{mMIMOoverview}. However, due to the drawbacks of OFDM transmission (e.g., high peak-to-average-power ratio (PAPR)), the use of single carrier in massive MIMO systems employing mm-wave bands, exhibiting sparsity both in angle and delay plane, was considered in \cite{SCvsOFDM,SCoptimality,mmWaveSC_Hybrid_mMIMO,GMGuvensen,Anil}. In these studies, the mitigation of ISI via reduced complexity beamspace processing (rather than temporal processing) motivates the use of SC in spatially correlated wideband massive MIMO systems \cite{mmWaveSC_Hybrid_mMIMO,GMGuvensen,Anil}.
In this study, we integrated NOMA with single-carrier uplink transmission as ISI is mitigated. We propose a novel reduced complexity receiver architecture for uplink MUSA based on our model. It surpasses the performance of conventional MPA based SCMA receiver when the number of RF chains is lower than the number of active users. The proposed receiver is based on both SIC and parallel interference cancellation (PIC) which has complexity in the order of the conventional MUSA receiver in \cite{MUSAforIOT}. The performance gap between the proposed and the state-of-the-art MMSE-SIC receiver is remarkable especially for higher loading factors and limited number of RF chains.

Performance analysis based on achievable information rate (AIR) via the mismatched decoding capacity \cite{Lapidoth}, where the suboptimality of practical receiver architectures and NOMA schemes are taken into account, is carried out. With the help of this analysis, for a given practical NOMA scheme (such as MUSA and SCMA) and number of RF chains, an optimal compromise between total spectral efficiency and number of active users (overloading factor) can be obtained. That is to say, it can be used as a design guideline to determine the minimum number of RF chain necessary to sustain a desired loading.

In the light of above contributions, the developed framework and proposed receiver architecture can support the use of MUSA in massive MIMO systems as a promising 5G modulation scheme. 

Part of this paper has appeared as a conference paper in \cite{bayraktar}. In this paper, detailed formulation of the proposed algorithm is provided and derivations for calculating mismatched decoding capacity is studied as an extension to that conference paper. Furthermore, highly overloaded scenarios are studied extensively, where single-carrier uplink MUSA is considered with the proposed algorithm. 

\subsubsection*{Notation}

Scalars, column vectors and matrices are denoted by lowercase (e.g., $ x $), lower-case boldface (e.g., $ \textbf{x} $) and uppercase boldface (e.g., $ \textbf{X} $) letters, respectively. $ \textbf{X}^T $, $ \textbf{X}^H $ and $ \textbf{X}^{-1} $ represent the transpose, Hermitian and inverse of the matrix $ \textbf{X} $, respectively. $ [ \textbf{X} ]_{i,j} $ is the entry of matrix $ \textbf{X} $ at $ i^{th} $ row and $ j^{th} $ column. $ \textbf{I} $ is the identity matrix with appropriate size and $ \textbf{I}_{N} $ is the identity matrix with size $ N \times N $. $ \mathbb{E} \{ \cdot \} $ and $ \mathrm{Tr} \{ \cdot \} $ are the expectation and trace operators, respectively. $ \lVert \textbf{x} \rVert $ denotes the Euclidean norm of vector $ \textbf{x} $. $ \textrm{diag} \{ \cdot \} $ and $ \textrm{bdiag} \{ \cdot \} $ are used to construct diagonal and block diagonal matrices with given inputs, respectively. $ P ( \cdot ) $ represents the probability value and $ p ( \cdot ) $ is the probability density function. $ \delta_{ij} $ is the Kronecker delta function.

\section{System Model}

\subsection{User-grouping Based Hybrid Architecture}

In this paper, a two-stage beamforming based massive MIMO system at mm-wave frequencies is investigated in time division duplexing (TDD) mode. There is a single BS with $ M $ antennas where the number of single antenna users is $ K $. Users are distributed to $ G $ groups since hybrid beamforming architecture based on JSDM is adapted \cite{JSDM}. Total number of RF chains is denoted by $ D $. Group $ g $ has $ K_g $ users and $ D_g $ RF chains are allocated to this group. Digital processing of each group becomes independent reducing the complexity significantly if the analog beamformer, which suppresses inter-group interference, is designed properly. Predetermined user groups are used in this paper.

\subsection{Frequency Selective Channel Model}

Instantaneous CSI is assumed to be known perfectly at the BS. In practice, channel vectors in reduced dimension are learned via uplink training in TDD mode \cite{GMGuvensen,Anil}. Channel matrix of group $ g $ for $ l^{th} $ MPC is defined as

\vspace{-8mm}
\begin{equation}
\textbf{H}_l^{(g)} = \Big[ \textbf{h}_l^{(g_1)} \; \textbf{h}_l^{(g_2)} \cdots \: \textbf{h}_l^{(g_{K_g})} \Big]_{M \times K_g} \label{H_l}
\vspace{-3mm}
\end{equation}

\noindent
for $ l = 0,\dots,L-1 $, where total number of MPCs is $ L $. Columns of $ \textbf{H}_l^{(g)} $ are channel vectors of each user in group $ g $. We use channel covariance matrix (CCM) based mm-wave channel model described in \cite{JSDM}. According to this model, channel vectors satisfy the following property:

\vspace{-6mm}
\begin{equation}
\mathbb{E}\bigg\{\textbf{h}_l^{(g_{m})} \Big[\textbf{h}_{l'}^{({g'}_{m'})}\Big]^H \bigg\} = \textbf{R}_l^{(g_{m})} \delta_{g-g'} \delta_{m-m'} \delta_{l-l'}, \label{E(H_l)}
\vspace{-3mm}
\end{equation}

\noindent
where $ \textbf{R}_l^{(g_{m})} $ is the CCM of $ m^{th} $ user in group $ g $ for $ l^{th} $ MPC. CCM of a MPC is defined as

\vspace{-4mm}
\begin{equation}
\textbf{R}_l^{(g_m)} \triangleq \gamma^{(g_m)} \int_{\mu_l^{(g_m)} - \frac{\Delta_l^{(g_m)}}{2}}^{\mu_l^{(g_m)} + \frac{\Delta_l^{(g_m)}}{2}} \rho_l^{(g_m)}(\theta) \textbf{u}(\theta) \textbf{u}(\theta)^H d\theta, \label{R_l}
\vspace{-3mm}
\end{equation}

\noindent
where $ \sqrt{\gamma^{(g_m)}} $ is the channel gain of $ m^{th} $ user in group $ g $ satisfying $ \sum_{l=0}^{L-1} \mathrm{Tr} \big\{\textbf{R}_l^{(g_m)} \big\} =  \gamma^{(g_m)}$ relation. For $ l^{th} $ MPC of $ m^{th} $ user in group $ g $, angular power profile function, mean AoA and angular spread are denoted by $ \rho_l^{(g_m)}(\theta) $, $ \mu_l^{(g_m)} $ and $ \Delta_l^{(g_m)} $, respectively. Unit norm steering vector corresponding to azimuth angle $ \theta $ with size $ M \times 1 $ is 

\vspace{-8mm}
\begin{equation}
\textbf{u}(\theta) \triangleq \frac{1}{\sqrt{M}} \big[1 \; e^{j \pi sin(\theta)} \cdots \: e^{j (M-1) \pi sin(\theta)} \big]^T. \label{u_theta}
\vspace{-4mm}
\end{equation}

\noindent
In this paper, CCMs are assumed to be slowly varying matrices which are perfectly known at the BS. We assume a correlated Rayleigh channel model where channel vectors are defined as $ \textbf{h}_l^{(g_{m})} \sim \mathcal{CN} \big( 0,\textbf{R}_l^{(g_{m})} \big) $ which is used in \cite{JSDM,Bjornson1,Bjornson2}.

\section{Downlink/Uplink Transmission Schemes for Hybrid Architecture}

In this paper, code-domain NOMA schemes are integrated with a user-grouping based hybrid beamforming architecture. We should define the codeword structure and its relation to the signal model before introducing the hybrid beamforming structure. In our framework, both in multicarrier downlink and single-carrier uplink scenarios, NOMA methods are applied inside groups since inter-group interference is efficiently suppressed. Therefore, number of users served in group $ g $ is $ K_g $. Number of shared resources (i.e., codeword length) is denoted by $ N_c $. Transmitted codeword belonging to user $ m $ in group $ g $ is defined as

\vspace{-6mm}
\begin{equation}
\textbf{c}_k^{(g_m)} = \Big[ c_{I_k^{(g_m)}}^{(g_m)}[0] \; c_{I_k^{(g_m)}}^{(g_m)}[1] \cdots \: c_{I_k^{(g_m)}}^{(g_m)}[N_c-1] \Big]^T \label{c_k_scma}
\vspace{-3mm}
\end{equation}

\noindent
for $ k = 0,1,\dots, N_s-1 $, where $ k $ is the NOMA symbol index and $ N_s $ is the total number of symbols transmitted in a frame. Selected symbol index of user $ m $ in group $ g $ for $ k^{th} $ symbol is $ I_k^{(g_m)} \in [1, \dots, \mathcal{M} ] $ where $ \mathcal{M} $ is the alphabet size. Codeword of this user for symbol $ I_k^{(g_m)} $ is defined as $ \Big\{ c_{I_k^{(g_m)}}^{(g_m)}[i], i = 0,\dots,N_c-1 \Big\} $ and $ \textbf{c}_k^{(g_m)} $ is the vector representation of the codeword. Note that analog beamformer suppresses inter-group signals effectively in our framework. As a consequence, same NOMA codewords can be used for different groups reducing the codeword generation overhead. On the other hand, user data vector at $ n^{th} $ physical resource is defined as

\vspace{-6mm}
\begin{equation}
\textbf{b}_n = \Big[ \big[\textbf{b}_n^{(1)}\big]^H \; \big[\textbf{b}_n^{(2)}\big]^H \cdots \: \big[\textbf{b}_n^{(G)}\big]^H \Big]^H, \label{b_k}
\vspace{-3mm}
\end{equation}

\noindent
where $\textbf{b}_n^{(g)}$ is the data vector of $g^{th}$ group satisfying $ \mathbb{E} \big\{ \textbf{b}_n^{(g)} \big[ \textbf{b}_n^{(g)} \big]^H \big\} = \textbf{I}_{K_g} $. User data vectors are expressed as  $\textbf{b}_n^{(g)} = \Big[ b_n^{(g_1)} \; b_n^{(g_2)} \dots \: b_n^{(g_{K_g})} \Big]^T $ where $ b_n^{(g_m)} $ is the data of user $ m $ in group $ g $. Transmitted code definition can be related to user data as

\vspace{-6mm}
\begin{equation}
b_n^{(g_m)} = \sum_{k=0}^{N_s - 1} c_{I_k^{(g_m)}}^{(g_m)}[n - N_c k],  \label{b_n_to_c_k}
\vspace{-3mm}
\end{equation}

\noindent
where $ n $ is used as subcarrier index for downlink multicarrier transmission and time index for uplink single-carrier transmission. 

\subsection{A Generic NOMA Signal Model for Downlink Multicarrier Transmission}

In this subsection, a hybrid beamforming based multicarrier system with $ N $ subcarriers is considered at mm-wave frequencies. Received signal at group $ g $ is expressed as

\vspace{-6mm}
\begin{equation}
\textbf{y}_n^{(g)} = \sum_{l=0}^{L-1} \big[ \textbf{H}_l^{(g)} \big]^H \textbf{x}_{(n-l)_N} + \textbf{n}_n, \label{y_n_dl_OFDM}
\vspace{-3mm}
\end{equation}

\noindent
where $ \textbf{x}_n  $ is the transmitted signal and $ \textbf{n}_n  $ is the noise vector at time index $ n $. Elements of the noise vector are assumed to be independent and identically distributed zero-mean complex circularly symmetric white Gaussian random variables with variance $ N_0 $. Transmitted signal $ \textbf{x}_{n} $ is expressed as 

\vspace{-6mm}
\begin{equation}
\textbf{x}_n = \textbf{B} \sum_{k=0}^{N-1} \frac{1}{\sqrt{N}} \mathrm{bdiag}\Big\{\sqrt{c^{(g)}}\textbf{W}_k^{(g)}\Big\}_{g=1}^G \textbf{P}_k \textbf{b}_k e^{j \frac{2 \pi}{N} kn},\label{x_n_OFDM}
\vspace{-3mm}
\end{equation}

\noindent
where the analog beamformer matrix with size $ M \times D $ is $ \textbf{B} = \big[ \textbf{S}^{(1)} \; \textbf{S}^{(2)} \cdots \: \textbf{S}^{(G)} \big] $ which consist of analog beamformers of all groups. Analog beamformer of group $ g $ which rejects the interference coming from other groups is denoted by $ \textbf{S}^{(g)} $ whose size is $ M \times D_g $. Digital beamformer of $ g^{th} $ group at $ k^{th} $ subcarrier is $\textbf{W}_k^{(g)} $ for $ k = 0,1, \dots, N-1 $ and power scaling factor of this group is $ c^{(g)} $. Selection of these matrices and parameters are explained in analog and digital beamforming subsections. Pattern matrix, represented by $ \textbf{P}_k $, is a $ K \times K $ diagonal matrix where diagonal entries are binary numbers indicating if the $ g_m^{th} $ user is transmitting data at $ k^{th} $ subcarrier. Transmitted data vector at $ k^{th} $ subcarrier, denoted by $ \textbf{b}_k $, was introduced in \eqref{b_k}. 

Operations in \eqref{x_n_OFDM} need to be explained further. In this equation, inverse DFT of the digitally precoded signal is taken before applying analog beamformer where $ \frac{1}{\sqrt{N}} $ is the normalization constant. In multicarrier transmission, user data can be decoded in frequency domain. Thus, DFT of the received signal is taken at the receiver which is expressed as $ \textbf{r}_k^{(g)} = \sum_{n=0}^{N-1} \frac{1}{\sqrt{N}} \textbf{y}_n^{(g)} e^{-j \frac{2 \pi}{N} kn} $. If $ \textbf{y}_n^{(g)} $ in \eqref{y_n_dl_OFDM} is substituted in this equation, DFT of the received signal becomes

\vspace{-6mm}
\begin{equation}\label{r_k_2}
\textbf{r}_k^{(g)} = \sum_{g' = 1}^{G} \sqrt{c^{(g')}} \Lambda_k^{(g',g)} \textbf{W}_k^{(g')}\textbf{b}_k^{(g')}  + \boldsymbol{\eta}_k,
\vspace{-3mm}
\end{equation}

\noindent
where $ \boldsymbol{\eta}_k $ is the DFT of the time domain noise vector and $ \Lambda_k^{(g',g)} $ is the effective channel in frequency domain that is expressed as $ \Lambda_k^{(g',g)} = \sum_{l=0}^{N-1} \big[ \textbf{H}_{eff,l}^{(g',g)} \big]^H e^{-j \frac{2 \pi}{N} lk} $. In other words, it is the DFT of complex conjugate of the reduced dimension effective channel matrices which are defined as $ \textbf{H}_{eff,l}^{(g',g) } \triangleq \big[ \textbf{S}^{(g')} \big]^H \textbf{H}_l^{(g)} $. Expression in \eqref{r_k_2} is rewritten as

\vspace{-6mm}
\begin{equation}\label{r_k_3}
\textbf{r}_k^{(g)} = \underbrace{\boldsymbol{\Phi}_k^{(g,g)}\textbf{b}_k^{(g)}}_{\text{Intra-Group Terms}} + \underbrace{\sum_{g' \neq g} \sqrt{c^{(g')}} \Lambda_k^{(g',g)} \textbf{W}_k^{(g')}\textbf{b}_k^{(g')} + \boldsymbol{\eta}_k}_{\boldsymbol{\xi}_k^{(g)} \text{: Inter-Group Interference + Noise Terms }},
\vspace{-3mm}
\end{equation}

\noindent
where intra-group and inter-group interference terms are separated since inter-group interference is assumed to be below noise level thanks to the analog beamformer. Channel gain matrix in \eqref{r_k_3} which includes only intra-group channels is defined as

\vspace{-8mm}
\begin{equation}
\boldsymbol{\Phi}_k^{(g,g)} = \sqrt{c^{(g)}} \Lambda_k^{(g,g)} \textbf{W}_k^{(g)}, \label{phi_k}
\vspace{-4mm}
\end{equation}

\noindent
whereas effective noise vector $ \boldsymbol{\xi}_k^{(g)} $ includes inter-group interference and noise terms. We can define the received symbol for a user at $ k^{th} $ subcarrier as 

\vspace{-6mm}
\begin{equation}
r_k^{(g_m)} = \big(\textbf{e}_m^{(g)}\big)^T \textbf{r}_k^{(g)} = \sum_{m'=1}^{K_g} \beta_k^{(g_m,g_{m'})} b_k^{(g_{m'})} + \xi_k^{(g_m)}, \label{r_k_gm}
\vspace{-3mm}
\end{equation}

\noindent
where $ \textbf{e}_m^{(g)} $ is a vector with length $ K_g $ whose all entries are zero except the $ m^{th} $ entry. This vector selects the correct symbol of each user in a group. Intra-group interference gain of user $ g_{m'} $ on user $ g_m $ in group $ g $ is given as $ \beta_k^{(g_m,g_{m'})} \triangleq \big[ \boldsymbol{\Phi}_k^{(g,g)} \big]_{(m,m')} $.

\subsubsection{Analog Beamformer Design}

We mainly focus on intra-group processing via different NOMA schemes. Thus, the suppression of inter-group interference is crucial with the help of analog beamformer which is constructed by using CCMs. Received vector with size $ M \times 1 $ at the BS during uplink transmission is expressed as

\vspace{-7mm}
\begin{equation}
\textbf{y}_n = \sum_{g=1}^{G} \sum_{l=0}^{L-1} \textbf{H}_l^{(g)} \textbf{x}_{n-l}^{(g)} + \textbf{n}_n, \label{y_n_ul}
\vspace{-4mm}
\end{equation}

\noindent
where $ \textbf{x}_n^{(g)} $ is the transmitted signal of group $ g $ at time index $ n $. Transmitted signal for each group can be defined in terms of user data as $ \textbf{x}_n^{(g)} \triangleq \sqrt{\frac{E_s^{(g)}}{K_g}} \textbf{b}_n^{(g)} $ where $ \textbf{b}_n^{(g)} $ is the user data vector of group $ g $ at time index $ n $ which was introduced in \eqref{b_k}. Total transmit energy per signaling interval of group $ g $ is  $ E_s^{(g)} $. Transmitted signals are created with a constraint given as $ \mathbb{E}\Big\{\textbf{x}_n^{(g)} \big[\textbf{x}_{n'}^{(g')}\big]^H \Big\} = \frac{E_s^{(g)}}{K_g} \textbf{I}_{K_g} \delta_{g-g'} \delta_{n-n'} $. With this constraint, covariance matrix of the received signal can be expressed as

\vspace{-6mm}
\begin{equation}
\textbf{R}_{\textbf{y}} = \mathbb{E}\Big\{\textbf{y}_n \big[\textbf{y}_{n}\big]^H \Big\} =  \sum_{g=1}^{G} \frac{E_s^{(g)}}{K_g} \sum_{l=0}^{L-1} \sum_{m=1}^{K_g}  \textbf{R}_l^{(g_{m})}  +  N_0 \textbf{I}_M, \label{R_y}
\vspace{-3mm}
\end{equation}

\noindent
where $ \textbf{R}_{\textbf{y}} $ includes CCMs of all users and their MPCs. 
While constructing analog beamformer, we adopt nearly optimal solution in \cite{GMGuvensen} where optimality is shown with respect to several criteria. In the light of this solution, we define $ \textbf{R}_{sum}^{(g)} = \sum_{l=0}^{L-1} \sum_{m=1}^{K_g} \textbf{R}_l^{(g_{m})} $ and solve the following generalized eigenvalue problem.

\vspace{-9mm}
\begin{equation}
\textbf{S}^{(g)} = \mathrm{eigs}(\textbf{R}_{sum}^{(g)},\textbf{R}_y,D_g). \label{AO-GEB}
\vspace{-6mm}
\end{equation}

\noindent
Columns of $ \textbf{S}^{(g)} $ are the most dominant $ D_g $ eigenvectors of the generalized eigenvalue problem defined above. This type of beamformer is suitable for multicarrier transmission since delay components are combined during DFT operation. 
This can be shown by finding the correlation matrix of the effective channel in frequency domain:

\vspace{-7mm}
\begin{equation}\label{E(lambda)}
\begin{aligned}
\textbf{R}_{\Lambda_k^{(g,g)}} & = \mathbb{E}\bigg\{\Lambda_k^{(g,g)} \Big[\Lambda_k^{(g,g)}\Big]^H \bigg\} = \big[ \textbf{S}^{(g)} \big]^H  \textbf{R}_{sum}^{(g)} \textbf{S}^{(g)}
\end{aligned}.
\vspace{-4mm}
\end{equation}

\noindent
It can be seen that correlation matrices of MPCs of users in group $ g $ are combined in the correlation matrix of the effective channel in frequency domain. Thus, using $ \textbf{R}_{sum}^{(g)} $ for the generalized eigenvalue problem is required when multicarrier transmission is utilized.

\subsubsection{Digital Beamformer Design}

In this paper, channel matched filter (CMF) type and CMF with spatially regularized zero-forcing (ZF) type beamformers are considered. These beamformers are defined as

\begin{equation}\label{W_k}
\textbf{W}_k^{(g)} \! \triangleq \!
\begin{dcases}
    \Big[ \Lambda_k^{(g,g)} \Big]^H,& \text{CMF}\\
    \Big[ \Lambda_k^{(g,g)} \Big]^H \bigg( \! \Lambda_k^{(g,g)} \Big[ \Lambda_k^{(g,g)} \Big]^H \! + \! \frac{K_g N_0}{E_s^{(g)}} \textbf{I}_{K_g} \! \bigg)^{-1} \!,              & \text{ZF}
\end{dcases},
\vspace{-3mm}
\end{equation}

\noindent
where $ \frac{K_g N_0}{E_s^{(g)}} \textbf{I}_{K_g} $ is the diagonal loading. Power scaling factors of digital beamformers are calculated based on a predetermined energy constraint. Transmit energy per subcarrier of the downlink system is defined as $ E_s = \sum_{g = 1}^{G} E_s^{(g)} = \mathbb{E} \big\{ \lVert \textbf{x}_n \rVert^2 \big\} $ where $ E_s^{(g)} $ is the transmit energy per subcarrier of group $ g $. If \eqref{x_n_OFDM} is substituted into $ E_s $ definition and block diagonal structure is used, transmit energy per subcarrier of group $ g $ is expressed as $ E_s^{(g)} = c^{(g)}P^{(g)} $. $ P^{(g)} $ is found as

\vspace{-6mm}
\begin{equation}\label{P_g}
P^{(g)} = \frac{1}{N} \sum_{k=0}^{N-1} \mathrm{Tr} \bigg\{  \textbf{S}^{(g)} \mathbb{E} \Big\{ \textbf{W}_k^{(g)} \textbf{P}_k^{(g)} \big[ \textbf{W}_k^{(g)} \big]^H \Big\} \big[ \textbf{S}^{(g)} \bigg]^H \bigg\},
\vspace{-3mm}
\end{equation}

\noindent
where $ \textbf{P}_k^{(g)} $ is the pattern matrix of group $ g $ at $ k^{th} $ subcarrier. Power scaling factor of group $ g $ is defined as $ c^{(g)} = E_s^{(g)} \big/ P^{(g)} $ that can be calculated by using Monte Carlo simulations.

\subsection{A Generic NOMA Signal Model for Uplink Single-Carrier Transmission}

\begin{figure}[h]
\centerline{\includegraphics[width=1\textwidth] 
{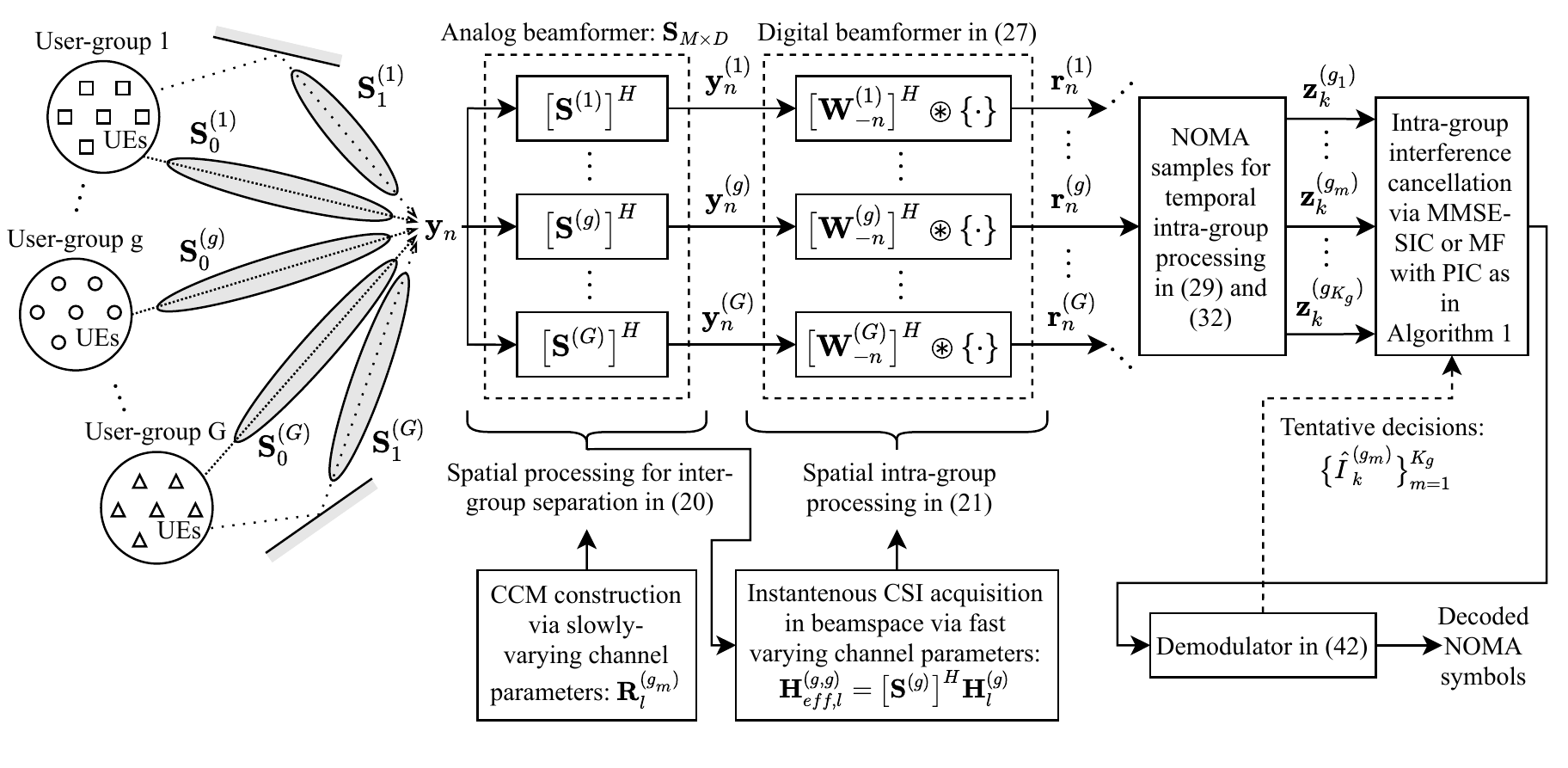}}
\caption{Block diagram of PIC-aided MUSA receiver with uplink single-carrier transmission}
\label{BlockDiagram}
\end{figure}

In this section, single-carrier model is employed for uplink transmission. Overall block diagram of the proposed MUSA receiver with uplink single-carrier transmission is given in Fig. \ref{BlockDiagram} and the blocks are explained in this and following sections. Received signal at the BS $ \textbf{y}_n $ was given in \eqref{y_n_ul} and transmitted signal of each group was defined as $ \textbf{x}_n^{(g)} \triangleq \sqrt{\frac{E_s^{(g)}}{K_g}} \textbf{b}_n^{(g)} $. At the BS, firstly analog beamformer of each group is applied to the received signal in order to separate the intended signal for group $ g $:

\vspace{-6mm}
\begin{equation}\label{y_n_g_ul}
\begin{aligned}
\textbf{y}_n^{(g)} & = \big[ \textbf{S}^{(g)} \big]^H \textbf{y}_n 
= \sum_{g'=1}^{G} \sum_{l=0}^{L-1} \big[ \textbf{S}^{(g)} \big]^H  \textbf{H}_l^{(g')} \textbf{x}_{n-l}^{(g')} + \big[ \textbf{S}^{(g)} \big]^H \textbf{n}_n \\
& = \underbrace{\sum_{l=0}^{L-1} \textbf{H}_{eff,l}^{(g,g)}  \textbf{x}_{n-l}^{(g)}}_{\text{Intra-Group Terms}} + \underbrace{ \sum_{g' \neq g} \sum_{l=0}^{L-1} \textbf{H}_{eff,l}^{(g,g')}  \textbf{x}_{n-l}^{(g')} + \big[ \textbf{S}^{(g)} \big]^H \textbf{n}_n}_{\boldsymbol{\eta}_n^{(g)} \text{: Inter-Group Interference + Noise Terms}}
\end{aligned},
\vspace{-3mm}
\end{equation}

\noindent
where inter-group interference and noise terms are gathered in $ \boldsymbol{\eta}_n^{(g)} $ since effect of interference terms are comparable with noise level in reduced dimension thanks to the analog beamformer. Recall that effective channel matrices in reduced dimension are defined as $ \textbf{H}_{eff,l}^{(g,g') } \triangleq \big[ \textbf{S}^{(g)} \big]^H \textbf{H}_l^{(g')} $. After the analog beamformer, digital beamformer is applied which can be shown as

\vspace{-6mm}
\begin{equation}
\textbf{r}_n^{(g)} = \sum_{l=0}^{L-1} \big[ \textbf{W}_l^{(g)} \big]^H \textbf{y}_{n+l}^{(g)}, \label{r_n_g_ul}
\vspace{-4mm}
\end{equation}

\noindent
where $ \textbf{W}_l^{(g)} $ is the digital beamformer of group $ g $ for $ l^{th} $ MPC. At the output of digital beamformers, intended signals of each user in group $ g $ are obtained as $ \textbf{r}_n^{(g)} $. If \eqref{y_n_g_ul} is substituted into \eqref{r_n_g_ul}, $ \textbf{r}_n^{(g)} $ becomes

\vspace{-8mm}
\begin{equation}
\textbf{r}_n^{(g)} = \underbrace{\boldsymbol{\Phi}_0^{(g,g)} \textbf{x}_{n}^{(g)}}_{\substack{\text{Intra-Group Terms} \\ \text{without ISI}}} + \underbrace{\sum_{{\substack{l=-(L-1) \\ l \neq 0}}}^{L-1} \boldsymbol{\Phi}_l^{(g,g)} \textbf{x}_{n+l}^{(g)} +\sum_{l=0}^{L-1} \big[ \textbf{W}_l^{(g)} \big]^H \boldsymbol{\eta}_{n+l}^{(g)}}_{\boldsymbol{\xi}_n^{(g)} \text{: ISI + Inter-Group Interference + Noise Terms}}, \label{r_n_g_ul_mod}
\vspace{-3mm}
\end{equation}

\noindent
where effective noise vector which is the summation of ISI, inter-group interference and noise terms is denoted by $ \boldsymbol{\xi}_n^{(g)} $. Channel gain matrices of each MPC of group $ g $ are defined as

\vspace{-6mm}
\begin{equation}
\boldsymbol{\Phi}_l^{(g,g)} = \sum_{l'=0}^{L-1} \big[ \textbf{W}_{l'}^{(g)} \big]^H \textbf{H}_{eff,l+l'}^{(g,g)} \label{Phi_l}
\end{equation}

\noindent
for $ l = -(L-1),-(L-2),\dots,(L-1) $. Equation \eqref{r_n_g_ul_mod} and \eqref{r_k_3} has similar forms. The only difference is that single-carrier transmission introduces ISI which is mitigated with the help of DFT operation in multicarrier transmission. Intended signal and ISI terms are separated in \eqref{r_n_g_ul_mod}. With this definition, received symbol of each user can be obtained as

\vspace{-6mm}
\begin{equation}
r_n^{(g_m)} = \big(\textbf{e}_m^{(g)}\big)^T \textbf{r}_n^{(g)} = \sum_{m'=1}^{K_g} \beta_0^{(g_m,g_{m'})} b_n^{(g_{m'})} + \xi_n^{(g_m)}, \label{r_n_gm}
\vspace{-3mm}
\end{equation}

\noindent
where $ \textbf{e}_m^{(g)} $ is the same vector as in the multicarrier case. Similarly, intra-group interference of user $ g_{m'} $ on user $ g_m $ is defined as $ \beta_l^{(g_m,g_{m'})} \triangleq \sqrt{\frac{E_s^{(g)}}{K_g}} \big[ \boldsymbol{\Phi}_l^{(g,g)} \big]_{(m,m')} $.

\subsubsection{Analog Beamformer Design}

Analog beamformer defined in \eqref{AO-GEB} is not suitable for single-carrier transmission since ISI due to MPCs are observed. In this case, $ d_l^{(g)} $ RF chains are allocated to $ l^{th} $ MPC in group $ g $ in order to suppress not only inter-group interference but also interference of other MPCs. Detailed explanation of this beamformer is given in \cite{GMGuvensen}. Analog beamformer of group $ g $ is expressed as

\vspace{-9mm}
\begin{equation}
\textbf{S}^{(g)} = \big[ \textbf{S}_0^{(g)} \; \textbf{S}_1^{(g)} \cdots \: \textbf{S}_{L-1}^{(g)} \big]_{N \times D_g}, \label{S_g}
\vspace{-4mm}
\end{equation}

\noindent
where $ \textbf{S}_l^{(g)} $ is the beamforming matrix of $ l^{th} $ MPC of group $ g $ which is found by following generalized eigenvalue problem:

\vspace{-9mm}
\begin{equation}
\textbf{S}_l^{(g)} = \mathrm{eigs}(\textbf{R}_{l}^{(g)},\textbf{R}_y,d_l^{(g)}), \label{JAD-GEB}
\vspace{-4mm}
\end{equation}

\noindent
where sum of CCMs of group $ g $ for $ l^{th} $ MPC is defined as $ \textbf{R}_l^{(g)} = \sum_{m=1}^{K_g} \textbf{R}_l^{(g_{m})} $. Optimum selection of number of RF chains for each MPC can be satisfied by maximizing $ \prod_{l=0}^{L-1} \prod_{n=1}^{d_l^{(g)}} (1+\lambda_n^l) $ where $ \lambda_n^l $ is the $ n^{th} $ dominant eigenvalue of the generalized eigenvalue problem defined in \eqref{JAD-GEB}.

\subsubsection{Digital Beamformer Design}

CMF type and CMF with spatially regularized ZF type beamformers are considered as in uplink single-carrier transmission. These beamformers are defined as

\vspace{-6mm}
\begin{equation}\label{W_l}
\textbf{W}_l^{(g)} \! = \!
\begin{dcases}
    \textbf{H}_{eff,l}^{(g,g)},& \! \! \! \text{CMF}\\
    \textbf{H}_{eff,l}^{(g,g)} \Bigg( \! \sum_{l'=0}^{L-1} \big[ \textbf{H}_{eff,l'}^{(g,g)} \big]^H \textbf{H}_{eff,l'}^{(g,g)} \! + \! \dfrac{K_g N_0}{E_s^{(g)}} \textbf{I}_{K_g} \! \! \Bigg)^{\! \! -1} \! \! \! ,              & \! \! \! \text{ZF}
\end{dcases}.
\end{equation}

\subsubsection{Mitigation of ISI in Beamspace for Single-Carrier Transmission}

In this framework, ISI is mitigated with the help of analog beamformer and there is no need for equalization. Channel gain matrix after the channel matched filter $ \boldsymbol{\Phi}_l^{(g,g)} $ definition should be revisited in order to explain this fact. If CMF type digital beamformer is used, channel gain matrix can be written as

\vspace{-6mm}
\begin{equation}\label{Phi_l_1}
\boldsymbol{\Phi}_l^{(g,g)} = \sum_{l'=0}^{L-1} \big[ \textbf{H}_{eff,l'}^{(g,g)} \big]^H \textbf{H}_{eff,l+l'}^{(g,g)} \\
= \sum_{l'=0}^{L-1} \big[ \textbf{H}_{l'}^{(g)} \big]^H \textbf{S}^{(g)} \big[ \textbf{S}^{(g)} \big]^H \textbf{H}_{l+l'}^{(g)}.
\vspace{-3mm}
\end{equation}

\noindent
Note that joint angle-delay analog beamformer is used for single-carrier transmission. $ \textbf{S}^{(g)} $ is constructed by concatenating $ \textbf{S}_l^{(g)} $ matrices which reject channels of MPCs except the channels belonging to $ l^{th} $ MPC. If $ l \neq 0 $ in \eqref{Phi_l_1}, terms in the summation become zero matrices since eigenspaces of different MPCs are orthogonal. Consequently, $ \boldsymbol{\Phi}_l^{(g,g)} $ contains non-zero entries if $ l \neq 0 $ and ISI terms in \eqref{r_n_g_ul_mod} become zero vectors. Thus, ISI is mitigated in the beamspace and $ \xi_n^{(g_m)} $ in \eqref{r_n_gm} contains only inter-group interference and noise terms as in the multicarrier transmission framework. On the other hand, $ \boldsymbol{\Phi}_0^{(g,g)} $ combine power coming from all MPCs. Therefore, multipath diversity is not lost while ISI is mitigated \cite{GMGuvensen}.

\section{A Reduced Complexity NOMA Receiver Architecture}

\subsection{Equivalent MIMO Signal Model}

Receivers of code-domain NOMA schemes require received signals at $ N_c $ physical resources in order to detect the transmitted codeword. Thus, following equivalent signal vector is supplied to receivers for both multicarrier downlink and single-carrier uplink transmission:

\vspace{-6mm}
\begin{equation}
\textbf{z}_k^{(g_m)} = \Big[	r_{k N_c}^{(g_m)} \; r_{k N_c + 1}^{(g_m)} \cdots \: r_{k N_c+N_c-1}^{(g_m)}	\Big]^T \label{z_k}
\vspace{-3mm}
\end{equation}

\noindent
for $ k = 0,1,\dots, N_s $. This vector should be expressed in terms of intra-group interference gains and codewords defined in \eqref{c_k_scma} which can be satisfied by using the relation in \eqref{b_n_to_c_k}. Equivalent signal models are given separately for downlink and uplink modes since channel gains are defined differently.

\subsubsection{Downlink Model}

If multicarrier downlink scenario is considered, \eqref{r_k_gm} should be substituted in \eqref{z_k} which results in

\vspace{-6mm}
\begin{equation}
\textbf{z}_k^{(g_m)} = \sum_{m'=1}^{K_g} \mathrm{diag} \big\{ \beta_{k N_c+i}^{(g_m,g_{m'})} \big\}_{i=0}^{N_c-1} \textbf{c}_{k}^{(g_{m'})} + \textbf{n}_k^{(g_m)}, \label{z_k_modified_MC}
\vspace{-3mm}
\end{equation}

\noindent
where equivalent noise vector of user $ m $ in group $ g $ for $ k^{th} $ NOMA symbol is defined as 

\vspace{-6mm}
\begin{equation}
\textbf{n}_k^{(g_m)} = \big[ \xi_{kN_c}^{(g_m)} \; \xi_{kN_c + 1}^{(g_m)} \dots \xi_{kN_c + N_c - 1}^{(g_m)} \big]^T. \label{n_k_NOMA}
\vspace{-3mm}
\end{equation}

\subsubsection{Uplink Model}

Equivalent signal model $ \textbf{z}_k^{(g_m)} $ can be rewritten for single-carrier uplink scenario as

\vspace{-8mm}
\begin{equation}
\textbf{z}_k^{(g_m)} = \sum_{m'=1}^{K_g} \mathrm{diag} \big\{ \beta_0^{(g_m,g_{m'})} \big\}_{i=0}^{N_c-1} \textbf{c}_{k}^{(g_{m'})} + \textbf{n}_k^{(g_m)} \label{z_k_modified_SC}
\vspace{-3mm}
\end{equation}

\noindent
if \eqref{r_n_gm} is substituted in \eqref{z_k}. Note that $ \textbf{n}_k^{(g_m)} $ has the same structure as in \eqref{n_k_NOMA} where entries of this vector are taken from \eqref{r_n_g_ul_mod}. In addition, channel gains in \eqref{z_k_modified_MC} and \eqref{z_k_modified_SC} are defined differently. Equivalent signal models for both downlink and uplink scenarios are suitable for NOMA receivers that are defined for single-input single-output (SISO) systems. In this paper, SCMA and MUSA schemes are integrated with the proposed architecture.

\subsection{Exemplary NOMA Scheme: SCMA}

SCMA, one of the strongest candidates for next generation multiple access standards \cite{SurveyNOMA,SurveyNOMA2,SurveyNOMA3}, is integrated with the proposed hybrid-beamforming based model as a case study. SCMA symbols can be decoded by MPA which uses factor graph structure. In our framework, $ \textbf{z}_k^{(g_m)} $ which is the input of MPA for both uplink and downlink transmission is similar to the one defined for uplink transmission in \cite{vaezi}. Sparse codewords are $ \textbf{c}_{k}^{(g_{m'})} $ and channel gains of users are $ \big\{ \beta_{k N_c+i}^{(g_m,g_{m'})} \big\}_{i=0}^{N_c-1} $ and $ \beta_0^{(g_m,g_{m'})} $ for downlink multicarrier and uplink single-carrier transmission, respectively. Thus, the same MPA and factor graph solution can be used to decode NOMA symbols. Furthermore, different codebooks should be assigned to users in downlink scenarios where MIMO is not used \cite{SCMACodebookDesign}. In this framework, signals of different users experience different channel gains which means we can use same codebooks for different users in the same group in downlink transmission.

\subsection{Exemplary NOMA Scheme: MUSA}

MUSA is a code-domain NOMA scheme which is similar to CDMA where spreading codes are used. There are two important differences between them. Spreading codes are shorter and overloading is realized where the number of users is greater than the code length for MUSA. Transmitted code structure in \eqref{c_k_scma} can be used for MUSA as well. Different symbols of users have different codewords in SCMA, whereas each user has a distinct spreading sequence which spreads a modulated symbol (e.g., QAM symbol) in MUSA. Hence, we can define transmitted code in terms of spreading sequences and modulated symbols as

\vspace{-8mm}
\begin{equation}
\textbf{c}_k^{(g_m)} \triangleq \textbf{s}^{(g_m)} a_{I_k^{(g_m)}}, \label{c_k_musa}
\vspace{-4mm}
\end{equation}

\noindent
where modulated symbol of user $ m $ in group $ g $ for $ k^{th} $ NOMA symbol is $ a_{I_k^{(g_m)}} $ and spreading code of this user is $ \textbf{s}^{(g_m)} = \big[s_0^{g_m} \cdots \: s_{N_c}^{g_m} \big]^T $. Classical MUSA receiver, SIC based MMSE, can be used for both downlink and uplink models.

\subsubsection{Downlink MUSA Model}

In downlink multicarrier scenario, equivalent received vector is defined in \eqref{z_k_modified_MC}. If \eqref{c_k_musa} is substituted in this vector, $ \textbf{z}_k^{(g_m)} $ becomes

\vspace{-6mm}
\begin{equation}
\begin{aligned}
\textbf{z}_k^{(g_m)} & = \sum_{m'=1}^{K_g} \mathrm{diag} \big\{ \beta_{k N_c+i}^{(g_m,g_{m'})} \big\}_{i=0}^{N_c-1} \textbf{s}^{(g_{m'})} a_{I_k^{(g_{m'})}} + \textbf{n}_k^{(g_m)} \\
& = \textbf{H}_{eq,k}^{(g_m)} \textbf{a}_k^{(g)} + \textbf{n}_k^{(g_m)}
\end{aligned}, \label{z_k_modified_MC_musa}
\vspace{-3mm}
\end{equation}

\noindent
where data vector is $ \textbf{a}_k^{(g)} = [a_{I_k^{(g_{1})}}, \dots, a_{I_k^{(g_{K_g})}}]^T $ and equivalent channel matrix is $ \textbf{H}_{eq,k}^{(g_m)} = \Big[ \textbf{h}_{eq,k}^{(g_m,g_{1})} \cdots \: \textbf{h}_{eq,k}^{(g_m,g_{K_g})}  \Big]  $. Equivalent channel vector of user $ g_{m'} $ for the stream of user $ g_{m} $ is defined as $ \textbf{h}_{eq,k}^{(g_m,g_{m'})} = \mathrm{diag} \big\{ \beta_{k N_c+i}^{(g_m,g_{m'})} \big\}_{i=0}^{N_c-1} \textbf{s}^{(g_{m'})} $. Matrix representation introduced in \eqref{z_k_modified_MC_musa} is required for MMSE detector defined as 

\vspace{-8mm}
\begin{equation}
\textbf{W}_{MMSE,k}^{(g_m)} =  \big[ \textbf{H}_{eq,k}^{(g_m)} \big]^H \Big( \textbf{H}_{eq,k}^{(g_m)} \big[ \textbf{H}_{eq,k}^{(g_m)}  \big]^H  + \textbf{R}_{\textbf{n}_k}^{(g_m)} \Big)^{-1}, \label{W_MMSE}
\vspace{-4mm}
\end{equation}

\noindent
where correlation matrix of the equivalent noise vector defined in \eqref{n_k_NOMA} is included in the MMSE detector. This correlation matrix is expressed as $ \textbf{R}_{\textbf{n}_k}^{(g_m)} = \mathbb{E} \Big\{ \textbf{n}_k^{(g_m)} \big[ \textbf{n}_k^{(g_m)} \big]^H \Big\} $. Entries of $ \textbf{n}_k^{(g_m)} $ are the $ m^{th} $ entries of effective noise vectors defined in \eqref{r_k_3} belonging to different subcarriers. Hence, it can be shown that

\vspace{-10mm}
\begin{equation}
\big[ \textbf{R}_{\textbf{n}_k}^{(g_m)} \big]_{(i,j)} = \big[ \textbf{R}_{\boldsymbol{\xi}}^{(g)} [i-j] \big]_{(m,m)}
\label{R_n_k}
\vspace{-4mm}
\end{equation}

\noindent
for $ i,j = 1,2,\dots,N_c $ and $ m = 1,2,\dots,K_g $, where correlation matrix of the effective noise vector is defined as $ \textbf{R}_{\boldsymbol{\xi}}^{(g)} [p] = \mathbb{E} \Big\{ \boldsymbol{\xi}_n^{(g)} \big[ \boldsymbol{\xi}_{n-p}^{(g)} \big]^H \Big\} $. If inter-group interference in \eqref{r_k_3} is ignored, $ \boldsymbol{\xi}_n^{(g)} $ includes only $ \boldsymbol{\eta}_n^{(g)}  $ which is the DFT of the time domain noise vector. Thus, it can be shown that $ \textbf{R}_{\textbf{n}_k}^{(g_m)} = N_0 \textbf{I}_{N_c} $ as noise is assumed to be independent both in time and among users with variance $ N_0 $.

MMSE detector defined in \eqref{W_MMSE} is used to find estimates of modulated symbols from the observations of user $ g_m $ as $ \hat{\textbf{a}}_k^{(g_m)} = \textbf{W}_{MMSE,k}^{(g_m)} \textbf{z}_k^{(g_m)} $. MMSE receiver can be improved by introducing SIC mechanism where users need to be ordered according to their signal-to-interference-plus-noise ratio (SINR) which is expressed as

\vspace{-6mm}
\begin{equation}
SINR_k^{(g_m,g_{m'})} =  \big[  \textbf{h}_{eq,k}^{(g_m,g_{m'})}  \big]^H \Bigg( \sum_{j \neq m'} \textbf{h}_{eq,k}^{(g_m,g_{j})}  \big[ \textbf{h}_{eq,k}^{(g_m,g_{j})} \big]^H + \textbf{R}_{\textbf{n}_k}^{(g_m)} \Bigg)^{-1} \textbf{h}_{eq,k}^{(g_m,g_{m'})},
\label{SINR_k}
\vspace{-3mm}
\end{equation}

\noindent
where $ SINR_k^{(g_m,g_{m'})} $ is the SINR of user $ g_{m'} $ after applying MMSE filter to the observations of user $ g_m $. After SINR ordering, MMSE estimate of the user with the highest SINR is found and this estimate is subtracted from the observation. Same MMSE process is applied to the remaining users until symbols of all users are detected. Details of the MMSE-SIC receiver for MUSA is discussed in \cite{NOMA_Comparison}.

\subsubsection{Uplink MUSA Model with Novel PIC-Aided Receiver}

Uplink transmission of MUSA has the same structure as in downlink scenario. A similar equivalent signal model that is defined for downlink transmission in \eqref{z_k_modified_MC_musa} can be obtained by using \eqref{z_k_modified_SC} and \eqref{c_k_musa} for uplink model. In this case, equivalent channel vector should be defined as $ \textbf{h}_{eq,k}^{(g_m,g_{m'})} = \mathrm{diag} \big\{ \beta_{0}^{(g_m,g_{m'})} \big\}_{i=0}^{N_c-1} \textbf{s}^{(g_{m'})} $. MMSE detector and SINR expressions are same as in previous part. Correlation matrix of equivalent noise vector can be found by using the equality in \eqref{R_n_k}. However, right-hand side of this equality is different from the one defined in previous part as effective noise vector $ \boldsymbol{\xi}_n^{(g)} $ is different for uplink single-carrier transmission. In order to calculate $ \textbf{R}_{\boldsymbol{\xi}}^{(g)} [p] $, correlation matrix of the noise after the beamforming $ \boldsymbol{\eta}_n^{(g)} $ should be calculated. If inter-group interference in \eqref{y_n_g_ul} is neglected, it can be expressed as

\vspace{-6mm}
\begin{equation}
\textbf{R}_{\boldsymbol{\eta}}^{(g)} [p] = \mathbb{E} \Big\{ \boldsymbol{\eta}_n^{(g)} \big[ \boldsymbol{\eta}_{n-p}^{(g)} \big]^H \Big\} = N_0 \big[ \textbf{S}^{(g)} \big]^H \textbf{S}^{(g)} \delta [p].
\label{R_eta}
\vspace{-3mm}
\end{equation}

\noindent
Using $ \textbf{R}_{\boldsymbol{\eta}}^{(g)} [p]  $ and ignoring ISI terms in \eqref{r_n_g_ul_mod} as they are mitigated in beamspace, $ \textbf{R}_{\boldsymbol{\xi}}^{(g)} [p] $ can be expressed as

\vspace{-6mm}
\begin{equation}
\textbf{R}_{\boldsymbol{\xi}}^{(g)} [p] = \mathbb{E} \Big\{ \boldsymbol{\xi}_n^{(g)} \big[ \boldsymbol{\xi}_{n-p}^{(g)} \big]^H \Big\} = N_0 \sum_{l=0}^{L-1} \big[ \textbf{W}_{l}^{(g)} \big]^H \big[ \textbf{S}^{(g)} \big]^H \textbf{S}^{(g)} \textbf{W}_{l+p}^{(g)}.
\label{R_ksi}
\vspace{-3mm}
\end{equation}

\noindent 
MMSE-SIC should be applied by using the correct $ \textbf{R}_{\textbf{n}_k}^{(g_m)} $ as described above so that the receiver works properly. Unlike conventional uplink systems, each user has its own stream in our framework where MMSE-SIC can be applied to each stream in group $ g $ resulting in $ K_g $ estimates for user $ g_m $. A novel receiver architecture is proposed for this situation. Initial SINRs are calculated before the MMSE-SIC process. After MMSE-SIC, decision for the symbol of a user is made by using the estimate of the stream where the user has the highest initial SINR in order to increase the reliability. Index of the stream which has the highest initial SINR for the symbol of user $ g_{m'} $ can be expressed as

\vspace{-6mm}
\begin{equation}
u = \underset{m}{\arg\max} \: SINR_k^{(g_m,g_{m'})}. \label{highest_SINR}
\vspace{-3mm}
\end{equation}

\noindent
This statement tells us that MMSE-SIC estimate obtained from $ \textbf{z}_k^{(g_u)} $ is used to detect the symbol of $ g_{m'} $. Index of the detected symbol is represented by $ \hat{I}_k^{(g_{m'})} $. Since detected symbol index of each user is available at the BS in uplink transmission, one can improve the performance by utilizing PIC method. For user $ g_{m} $, interference coming from other users in group $ g $ are subtracted from the observation $ \textbf{z}_k^{(g_m)} $ by using the symbols detected via MMSE-SIC receiver. After that, equivalent channel matched filter is applied in order to find the estimated symbol. PIC process can be expressed as  

\vspace{-6mm}
\begin{equation}
\hat{a}_k^{(g_{m})} = \frac{\big[ \textbf{h}_{eq,k}^{(g_{m},g_{m})} \big]^H}{ \big\lVert \textbf{h}_{eq,k}^{(g_{m},g_{m})} \big\lVert^2 } \Bigg[ \textbf{z}_k^{(g_m)} - \sum_{m' \neq m}\textbf{h}_{eq,k}^{(g_m,g_{m'})} a_{\hat{I}_k^{(g_{m'})}}  \Bigg], \label{PIC}
\vspace{-3mm}
\end{equation}

\noindent
where $ \hat{I}_k^{(g_{m'})} $ for $ m' \neq m $ is obtained from MMSE-SIC receiver based on $ \textbf{z}_k^{(g_u)} $ and $ \hat{a}_k^{(g_{m})} $ is the estimated symbol of user $ g_{m} $ which is used to detect the new symbol index $ \hat{I}_k^{(g_{m})} $ via minimum distance rule that is expressed as

\vspace{-10mm}
\begin{equation}
\hat{I}_k^{(g_{m})} = \underset{I_k^{(g_{m})} }{\arg\min} \: \Big| \: \hat{a}_k^{(g_{m})} - a_{I_k^{(g_{m})}} \: \Big|^2 \label{min_dist}
\vspace{-3mm}
\end{equation} 

\noindent
concluding the PIC operation. However, PIC can be repeated several times by using the detected symbols of previous PIC iterations. Proposed PIC-aided uplink receiver is described in Algorithm \ref{PIC-aided} where $ max $\textunderscore$ iter $ is the maximum number of PIC iterations.

\vspace{-1mm}
\setlength{\textfloatsep}{10pt}
\begin{algorithm}
\setstretch{1.25}
\caption{PIC-aided uplink receiver for MUSA}\label{PIC-aided}
\begin{algorithmic}[1]

\Require $ max $\textunderscore$ iter $, $ \big\{ \textbf{z}_k^{g_m} $, $ \textbf{h}_{eq,k}^{(g_{m},g_{m'})} \big\}_{m,m' = 1}^{K_g} $ for $ g \!=  \! 1, \dots, G $
\Ensure $ \big\{ \hat{I}_k^{(g_{m})} \big\}_{m = 1}^{K_g} $ for $ g \!=  \! 1, \dots, G $ 

	\vspace{1.5mm}
    \hrule
    \vspace{1.5mm}
    
	\For {$ g \leq G $}	 \Comment{MMSE-SIC}
			\For {$ m \leq K_g $}
				\For {$ m' \leq K_g $}
					\State Find initial $ SINR_k^{(g_m,g_{m'})} $ as in \eqref{SINR_k}		
				\EndFor
				\State Apply MMSE-SIC		
			\EndFor				
			\For {$ m' \leq K_g $}
				\State $ u = \underset{m}{\arg\max} \: SINR_k^{(g_m,g_{m'})} $
				\vspace{1mm}
				\State Select $ \hat{I}_k^{(g_{m'})} $ obtained from $ \textbf{z}_k^{(g_u)} $ via MMSE-SIC
			\EndFor

	\EndFor
	
\For {$ i \leq max $\textunderscore$ iter $} \Comment{PIC iterations}
	\For {$ g \leq G $}		
	\For {$ m \leq K_g $}	
		\State Apply PIC as in \eqref{PIC} by using $ \big\{ \hat{I}_k^{(g_{m'})} \big\}_{m' \neq m} $
		\State Find $ \hat{I}_k^{(g_{m})} $ as in \eqref{min_dist}
	\EndFor
	\EndFor
\EndFor

\Return $ \big\{ \hat{I}_k^{(g_{m})} \big\}_{m = 1}^{K_g} $ for $ g \!=  \! 1, \dots, G $

\end{algorithmic}
\end{algorithm}

\vspace{-4mm}
\subsection{Complexity}

Complexity of MPA based SCMA receiver is $ \mathcal{O}(T K_g \mathcal{M}^{v}) $ where $ T $ is the number of iterations and $ v $ is the number of users with non-zero codeword entry at a physical resource \cite{vaezi}. Complexity of MMSE expression given in \eqref{W_MMSE} is $ \mathcal{O} (N_c^3) $ due to the matrix inverse. Complexity of MMSE-SIC becomes $ \mathcal{O} (K_g N_c^3) $ as SIC requires the calculation of MMSE expression at each step. PIC-aided uplink receiver for MUSA has a complexity in the order of MMSE-SIC receiver since PIC iteration given in \eqref{PIC} has a lower complexity order. These are the complexity orders of the MPA based SCMA and PIC-aided MUSA receiver applied to the stream of a user given in \eqref{z_k_modified_SC}. In other words, these receivers are applied to streams of every user. Complexity of SCMA increases tremendously with the cardinality, whereas MUSA receiver is not affected by it. On the other hand, code length increases the complexity of MUSA receiver. It is important to note that overloading increases the complexity of MUSA receivers linearly while both number of users and $ v $ increases for SCMA which causes excessive growth in complexity. It can be seen that MUSA receivers have lower complexity when compared to SCMA receiver especially for overloaded scenarios. Furthermore, MUSA has a more flexible structure as the number of users and spreading code length can be changed freely, whereas these parameters are fixed for SCMA.

\section{Performance Measures in terms of Achievable Information Rate (AIR)}

Codeword length and overloading factor are parameters that can be varied in NOMA schemes. Changing these parameters affects the bit-error rate (BER) which is a good measure of performance. However, comparing only BER performance for different NOMA parameters may not be adequate since rate per user depends on codeword length for fixed overloading factor and total rate depends on overloading factor for fixed codeword length. Hence, both BER and rate comparison are required for a comprehensive performance analysis. Achievable rate for a communication system with equally likely input symbols is defined as

\vspace{-6mm}
\begin{equation}
\mathrm{AIR} = log_2 \mathcal{M} - \mathbb{E}_{X,Y,h} \Bigg\{ log_2 \Bigg( \frac{\sum_{x} p(Y| x,h)}{p(Y | X,h)} \Bigg) \Bigg\}, \label{rate}
\vspace{-3mm}
\end{equation}

\noindent
where $ p(y|x,h) $ is the probability density function of output value $ y $ for given input symbol $ x $ and channel condition is denoted by $ h $ \cite{Ungerboeck}. Expectation is taken over different realizations of input and output values and also over varying channel conditions. Equation \eqref{rate} can be used to calculate AIR if conditional probability densities are exactly known. In our framework, even if we know the channel perfectly, inter-group interference which is neglected for both multicarrier and single-carrier transmission exists in addition to Gaussian noise. Moreover, NOMA receivers are suboptimal which causes mismatch. In the literature, this kind of decoders are called mismatched decoders. In \cite{Lapidoth}, a lower bound is proposed for \eqref{rate} by using mismatched probability density functions and it is proved that any probability density function other than the original one gives a lower achievable information rate. This lower bound can be given as

\vspace{-6mm}
\begin{equation}
\mathrm{AIR}' = log_2 \mathcal{M}  - \mathbb{E}_{X,Y,h} \Bigg\{ log_2 \Bigg( \frac{\sum_{x} \hat{p}(Y| x,h)}{\hat{p}(Y | X,h)} \Bigg) \Bigg\}, \label{rate2}
\vspace{-3mm}
\end{equation}

\noindent
where $ \hat{p}(y|x,h) $ is a mismatched probability density function that needs to be calculated at the output of NOMA receivers. It is important to point out that AIR expression should be divided by $ N_c $ for NOMA schemes since a NOMA symbol is transmitted by using $ N_c $ physical resources.

\vspace{-3mm}
\subsection{Decoding Capacity for Uplink SCMA}

SCMA uses a MPA based receiver which calculates mismatched probabilities of symbols given output values (i.e, $ \hat{P}(x|y,h) $) at its output. These probabilities can be used instead of $ \hat{P}(y|x,h) $ if symbols are equally likely. Lower bound on rate expression in \eqref{rate2} can be redefined as

\vspace{-6mm}
\begin{equation}
\mathrm{AIR}' = log_2 \mathcal{M}  - \mathbb{E}_{X,Y,h} \Bigg\{ log_2 \Bigg( \frac{\sum_{x} \hat{p}(x| Y,h)}{\hat{p}(X | Y,h)} \Bigg) \Bigg\}, \label{rate_SCMA}
\vspace{-3mm}
\end{equation}

\noindent
where $ \hat{p}(x| y,h) $ is the mismatched probability density function of input symbol given the output signal. Note that MPA does not provide probability density functions, instead it gives probability values which can be used for calculating AIR given in \eqref{rate_SCMA} via Monte Carlo simulations.

\subsection{Decoding Capacity for Uplink MUSA}

AIR is calculated for the outputs of PIC iterations given in \eqref{PIC} with two assumptions. The first one is that interference cancellation is assumed to be perfect. If this is the case, after interference cancellation \eqref{PIC} can be rewritten as

\vspace{-6mm}
\begin{equation}
\hat{a}_k^{(g_{m})} = a_{I_k^{(g_{m})}} + \frac{\big[ \textbf{h}_{eq,k}^{(g_{m},g_{m})} \big]^H \textbf{n}_k^{(g_m)}}{ \big\lVert \textbf{h}_{eq,k}^{(g_{m},g_{m})} \big\lVert^2 }, \label{PIC2}
\vspace{-3mm}
\end{equation}

\noindent
where the second term on the right-hand side is taken as noise and its variance is calculated as

\vspace{-4mm}
\begin{equation}
{N_0}' = \mathbb{E} \Bigg\{ \frac{\big[ \textbf{h}_{eq,k}^{(g_{m},g_{m})} \big]^H \textbf{n}_k^{(g_m)} \big[ \textbf{n}_k^{(g_m)} \big]^H \textbf{h}_{eq,k}^{(g_{m},g_{m})}}{ \big\lVert \textbf{h}_{eq,k}^{(g_{m},g_{m})} \big\lVert^4 } \Bigg\} = \frac{\big[ \textbf{h}_{eq,k}^{(g_{m},g_{m})} \big]^H \textbf{R}_{\textbf{n}_k}^{(g_m)} \textbf{h}_{eq,k}^{(g_{m},g_{m})}}{ \big\lVert \textbf{h}_{eq,k}^{(g_{m},g_{m})} \big\lVert^4 }, \label{N0_mod}
\end{equation}

\noindent
where the method to calculate $ \textbf{R}_{\textbf{n}_k}^{(g_m)} $ was given in the uplink MUSA section. Furthermore, expression given in \eqref{PIC2} is assumed to have a complex Gaussian distribution with noise variance $ {N_0}' $. By using this assumption, conditional probability density function can be written as

\vspace{-6mm}
\begin{equation}
\hat{p} \Big( \hat{a}_k^{(g_{m})} \big| I_k^{(g_{m})},\textbf{h}_{eq,k}^{(g_{m},g_{m})} \Big) = \frac{1}{ \pi {N_0}'} exp \Bigg( - \frac{ \big| \hat{a}_k^{(g_{m})} - a_{I_k^{(g_{m})}} \big|^2 }{ {N_0}'} \Bigg), \label{PDF_MUSA}
\vspace{-3mm}
\end{equation}

\noindent
where $ \hat{a}_k^{(g_{m})} $, $ I_k^{(g_{m})} $ and $ \textbf{h}_{eq,k}^{(g_{m},g_{m})} $ are used as observation $ y $, input symbol $ x $ and channel condition $ h $, respectively, in the generic probability density function $ \hat{p}(y|x,h) $. AIR can be calculated via \eqref{rate2} by using Monte Carlo simulations where probabilities are calculated with the expression given in \eqref{PDF_MUSA}.

\section{Numerical Evaluation}
\vspace{-2mm}
\subsection{Case Study}

In this section, BER and AIR results of several simulations are provided. A predetermined MIMO setting is used where BS has a uniform linear array (ULA) with $ M = 100 $ antennas and users have single antennas. Antenna spacing of ULA is set to half the wavelength where the carrier frequency is 30 GHz. Users are distributed equally in 4 groups. Although total number of delays is $ L = 32 $, users do not have active MPCs at every delay. In the simulations, first group has 3 MPCs, whereas other groups have 2 MPCs. The scenario used for simulations is given in Table \ref{table1}. This table shows the active MPC index of groups and angular sector (AS) for mean AoA $ \mu_l^{(g_m)} $ related to each MPC. Mean AoA of each user in a group at a MPC is selected randomly from the given AS. Angular spread $ \Delta_l^{(g_m)} $ of all users is set to $ 3 \degree $ for every MPC. Angle-delay map of groups is given in Fig. \ref{PDP} where AS for mean AoA and angular spread are both taken into account. Power profile function $ \rho_l^{(g_m)}(\theta) $ is assumed to be uniform in the angular spread of the MPC and channel gain $ \sqrt{\gamma^{(g_m)}} $ is  set to 1 for every MPC of all users. Channels are generated randomly according to model \eqref{R_l} by using these parameters. This scenario is typical for mm-wave where we observe a sparse angle-delay profile. BER curves are obtained via Monte Carlo simulations where 100 different channel realizations are generated. Total transmit energy $ E_s $ and RF chains are distributed equally to groups. Three different multiple access methods are investigated: SCMA, MUSA and CMF with ZF only. CMF and CMF with ZF type beamformers are used for SCMA and MUSA, respectively. Number of MPA iterations is set to 10. For CMF with ZF only case, symbols of all users are transmitted at the same resource blocks. Alphabet size $ \mathcal{M} $ is selected as 4 (QPSK) for all schemes. Codebook in \cite{vaezi} is used for SCMA, whereas spreading code elements of MUSA are selected randomly from the set $ \{0,1,1+j,j,-1+j,-1,-1-j,-j,1-j\} $ \cite{MUSAforIOT}. Users in different groups use same codebooks and spreading codes.

\vspace{-8mm}
\begin{table}[ht]
\small
\begin{minipage}[b]{0.46\linewidth}
\centering
\caption{Angle-delay profile of groups}
\vspace{-10mm}
\begin{center}
\begin{tabular}{|c|c|c|}
\hline
Group      & MPC index  & AS for mean AoA $\big( \mu_l^{(g_m)} \big) $\\ \hline
\multirow{3}{*}{1} & 0         & {[}-1.25, 1.5{]}      \\ \cline{2-3} 
                   & 5         & {[}8.25, 9.75{]}      \\ \cline{2-3} 
                   & 11        & {[}20.25, 22{]}      \\ \hline
\multirow{2}{*}{2} & 3         & {[}25, 27.5{]}     \\ \cline{2-3} 
                   & 9         & {[}13.5, 15.75{]}      \\ \hline
\multirow{2}{*}{3} & 8         & {[}-8, -6{]}      \\ \cline{2-3} 
                   & 17        & {[}-14.75, -12.5{]}    \\ \hline
\multirow{2}{*}{4} & 20        & {[}-21.5, -19.5{]}  \\ \cline{2-3} 
                   & 29        & {[}-28, -26{]}       \\ \hline
\end{tabular}
\end{center}
\label{table1}
\end{minipage}\hfill
\begin{minipage}[b]{0.46\linewidth}
\centering
\includegraphics[width=1\linewidth]{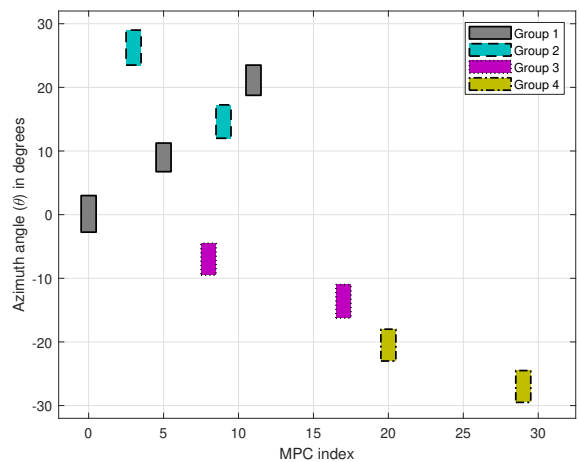}
\vspace{-8mm}
\captionof{figure}{Angle-delay map of groups}
\label{PDP}
\end{minipage}
\end{table}

\subsection{Simulation Results}

\subsubsection{SCMA and MUSA Comparison}

\begin{figure}[h]
\centering
\subfloat[Subfigure 1 list of figures text][Multicarrier downlink transmission]{
\includegraphics[width=0.46\textwidth]{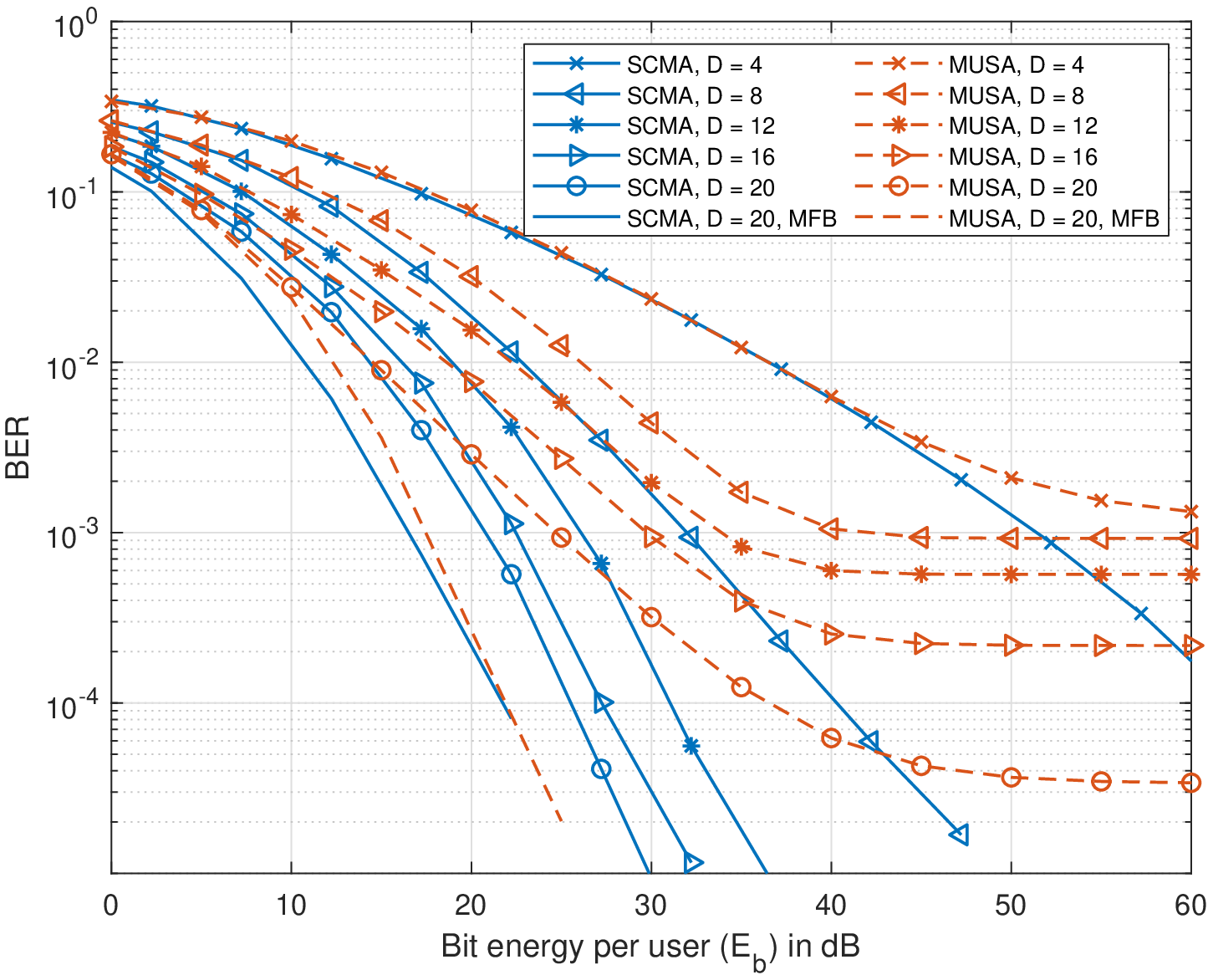}
\label{MC_MUSA_SCMA}}\hfill
\subfloat[Subfigure 2 list of figures text][Single-carrier uplink transmission]{
\includegraphics[width=0.46\textwidth]{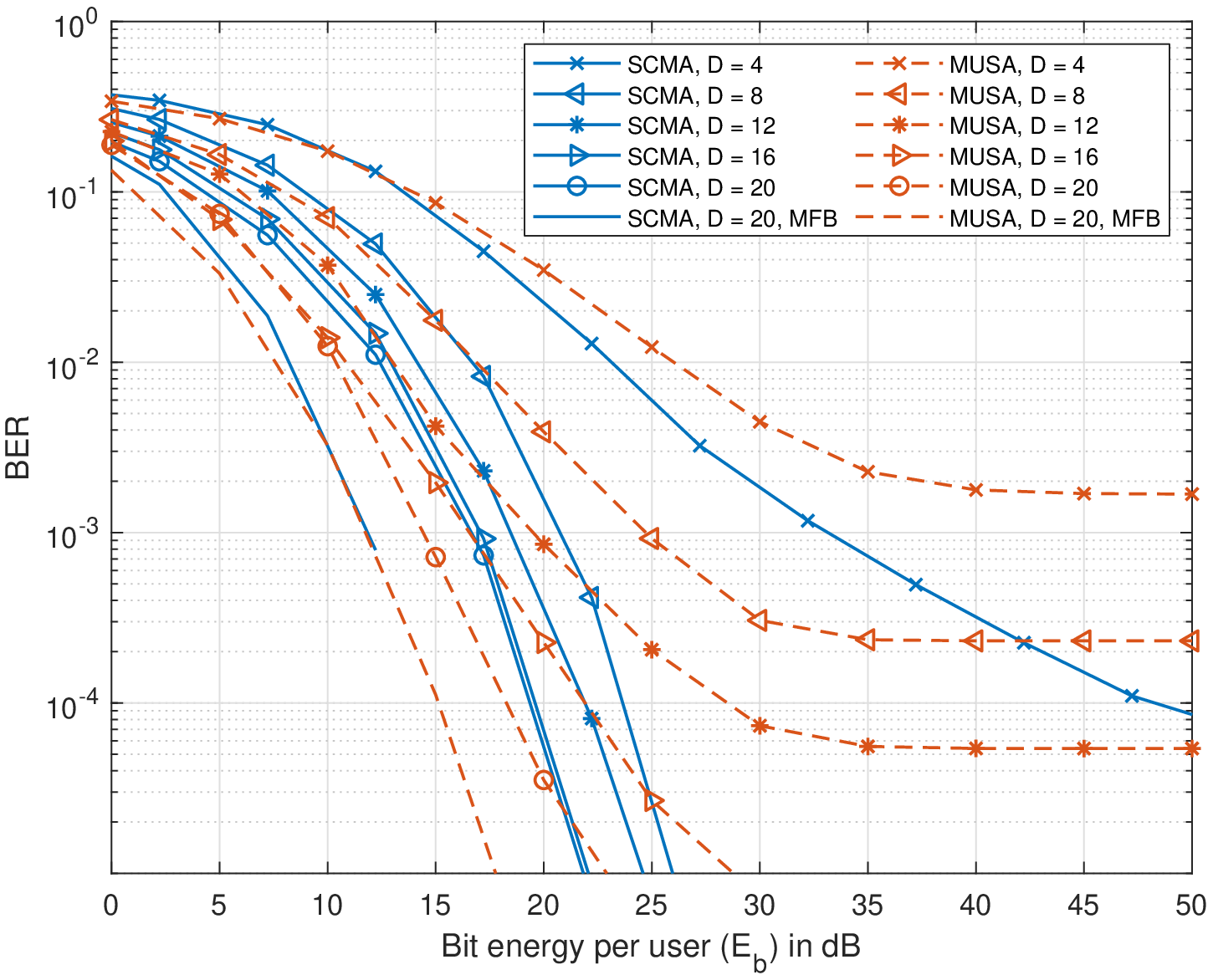}
\label{SC_MUSA_SCMA}}
\caption{Average BER vs. bit energy per user ($ E_b $) for MUSA and SCMA with $ K_g = 6 $}
\label{MUSA_SCMA_BER}
\end{figure}

Fig. \ref{MC_MUSA_SCMA} shows the BER comparison of MUSA and SCMA with $ K_g = 6 $ and $ N_c = 4 $ where multicarrier downlink is considered. It can be seen that performance of both MUSA and SCMA improve significantly as $ D $ increases. MUSA has an error floor at each case and SCMA clearly outperforms MUSA. On the other hand, PIC-aided uplink MUSA receiver improves the performance of MUSA tremendously as it can be seen from Fig. \ref{SC_MUSA_SCMA}. Error floor level of MUSA gets significantly lower with increasing $ D $ and it performs better than SCMA at low $ E_b $. Matched filter bound (MFB) is the case where intra-group interference is eliminated. MFB curves of SCMA and MUSA are close to each other and PIC makes MUSA perform close to this bound. In addition, we observe that there is no error floor for MFB curves showing the fact that inter-group interference is efficiently suppressed. 

\vspace{-6mm}
\vspace{-6mm}

\begin{figure}[h]
\centering
\subfloat[Subfigure 1 list of figures text][Multicarrier downlink transmission]{
\includegraphics[width=0.46\textwidth]{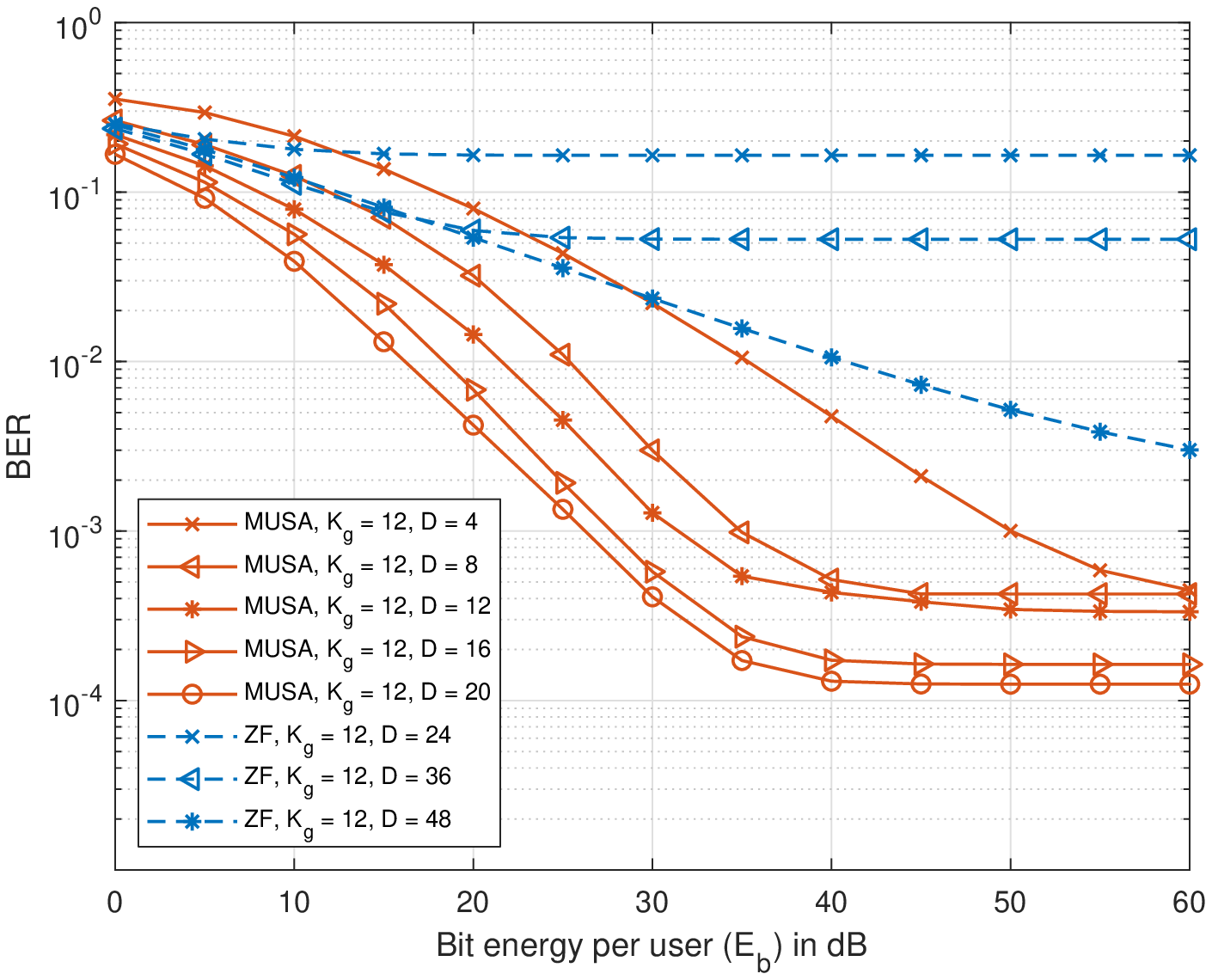}
\label{MC_MUSA_ZF}}\hfill
\subfloat[Subfigure 2 list of figures text][Single-carrier uplink transmission]{
\includegraphics[width=0.46\textwidth]{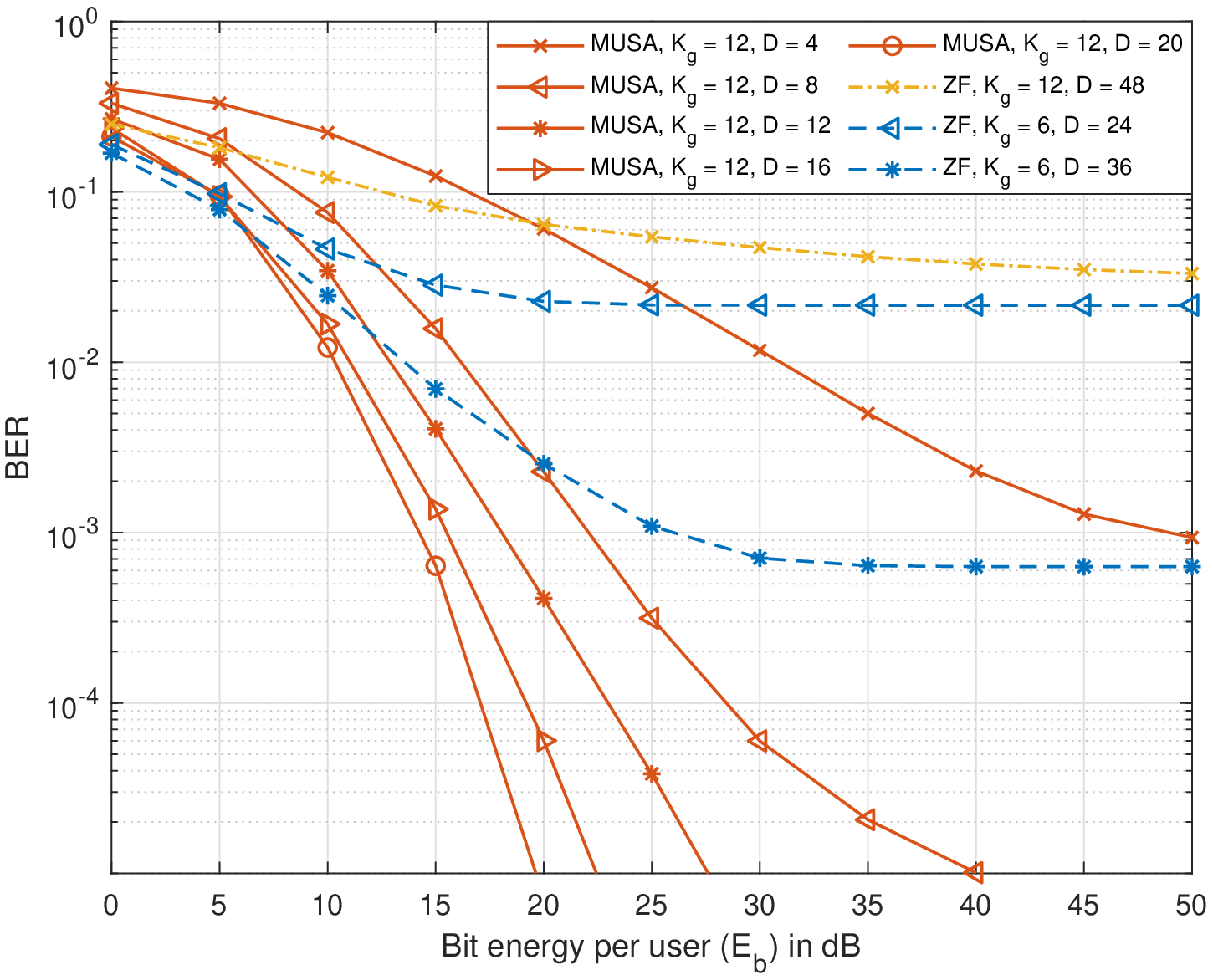}
\label{SC_MUSA_ZF}}
\caption{Average BER vs. bit energy per user ($ E_b $) for MUSA with $ N_c = 8 $ and ZF}
\label{MUSA_ZF_BER}
\end{figure}

Fig. \ref{MC_MUSA_ZF} compares BER performance of MUSA and CMF with ZF for multicarrier downlink transmission where $ K_g = 12 $ and $ N_c = 8 $. ZF performs poorly even when the number of RF chains is equal to the number of users, whereas MUSA has acceptable BER thanks to long spreading sequences. Fig. \ref{SC_MUSA_ZF} compares MUSA and CMF with ZF for single-carrier uplink transmission and it is seen that PIC-aided MUSA with $ K_g = 12 $ outperforms ZF. Error floor of MUSA disappears for $ D = 12 $, whereas ZF has an error floor even when $ K_g = 6 $ and $ D = 36 $. These results show that MUSA with long spreading sequences perform better than ZF when the number of closely located users is high.

\vspace{-6mm}
\vspace{-6mm}

\begin{figure}[h]
\centerline{\includegraphics[width=0.65\textwidth]{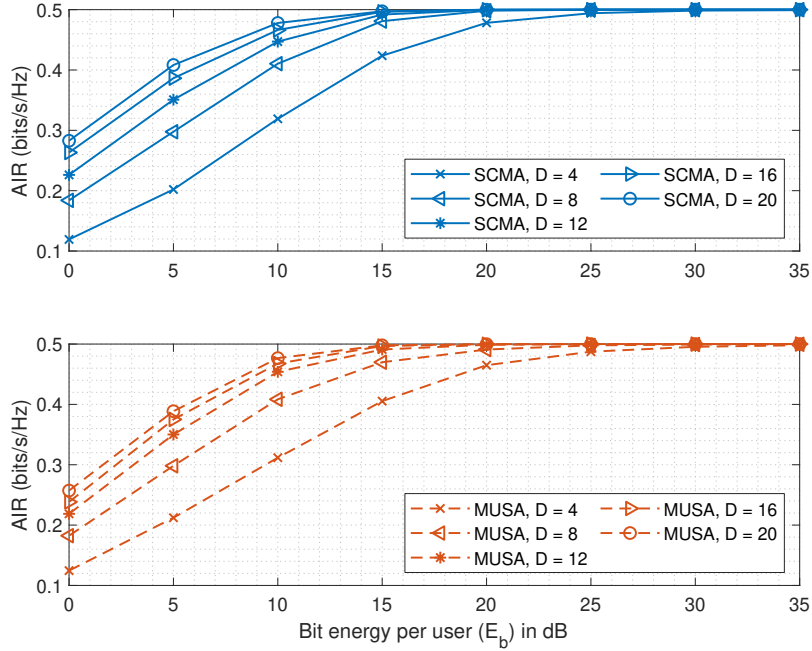}}
\vspace{-3mm}
\caption{Average AIR per user vs. bit energy per user ($ E_b $) for single-carrier uplink transmission with $ N_c = 4 $ and $ K_g = 6 $}
\label{MUSA_SCMA_AIR}
\vspace{3mm}
\end{figure}

Fig. \ref{MUSA_SCMA_AIR} compares AIR per user performances of SCMA and MUSA where $ N_c = 4 $ and $ K_g = 6 $ for uplink single-carrier transmission. It is seen that both NOMA schemes have very similar AIR curves. AIR performance improves significantly with increasing number of RF chains at low $ E_b $ for both schemes. In addition, AIR saturates at the maximum value as $ E_b $ increases for all $ D $ values. This means that NOMA schemes perform close to the maximum achievable rate even when each user group has a single RF chain if the bit energy is high enough.

\subsubsection{MUSA with High Loading Factors}

\begin{figure}[h]
\centering
\subfloat[Subfigure 1 list of figures text][Average BER vs. overloading]{
\includegraphics[width=0.46\textwidth]{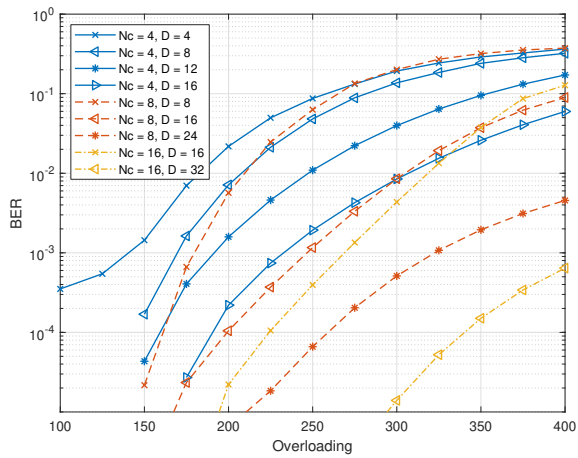}
\label{MUSA_Nc_D_BER}}\hfill
\subfloat[Subfigure 2 list of figures text][Average AIR per group vs. overloading]{
\includegraphics[width=0.46\textwidth]{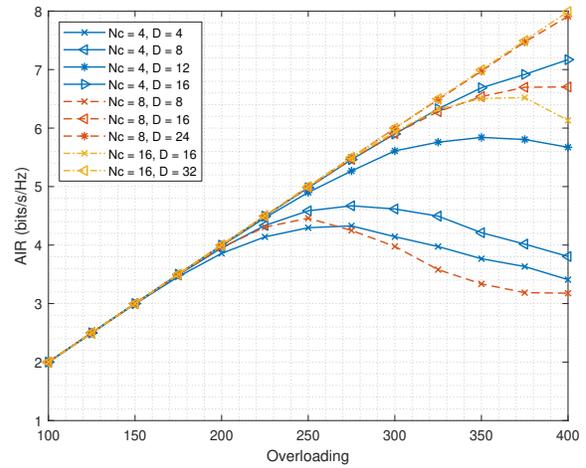}
\label{MUSA_Nc_D_AIR}}
\caption{Average BER and average AIR per group vs. overloading for uplink MUSA at $ E_b = $ 40 dB with different code lengths and number of RF chains}
\label{MUSA_Nc_D}
\end{figure}

In this section, flexible structure of MUSA is exploited and performance of PIC-aided uplink MUSA receiver is investigated. It is important to note that when code length increases, number of users in a group also increases at a fixed overloading. Therefore, it is meaningful to use more RF chains as code length increases so that the number of RF chains per user remains constant. Fig. \ref{MUSA_Nc_D_BER} shows the average BER vs. loading for various code length and number of RF chains at $ E_b = 40 $ dB. It can be seen that overloading decreases the BER performance as expected. Furthermore, using long spreading codes improves the performance significantly while keeping number of RF chains per user constant. Even though BER analysis provides some insights about the performance of MUSA for different $ N_c $, $ D $ and overloading, it is not possible to decide which overloading should be used for fixed $ N_c $ and $ D $ values. Therefore, average AIR per group curves are provided in Fig. \ref{MUSA_Nc_D_AIR}. It is seen that for low $ D $ values there is an overloading value that yields maximum rate. On the other hand, AIR curve stays linear when $ N_c = 16 $ and $ D = 32 $ which shows that AIR is at the maximum level even when the overloading is $ 400 \% $. This result shows that long spreading codes are required for highly dense networks. 

\begin{figure}[h]
\centering
\subfloat[Subfigure 1 list of figures text][Average BER vs. loading at $ E_b = $ 40 dB]{
\includegraphics[width=0.46\textwidth]{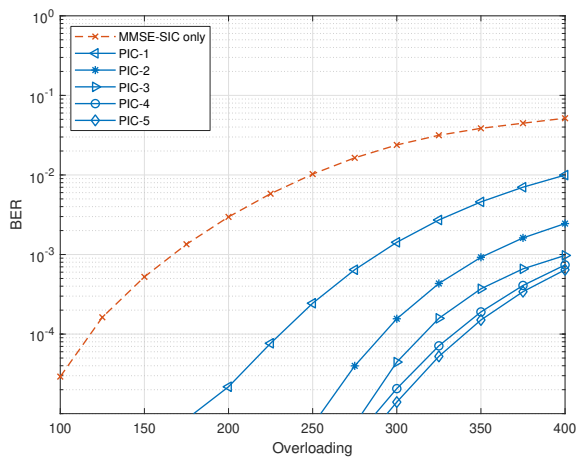}
\label{MUSA_iter_Nc}}\hfill
\subfloat[Subfigure 2 list of figures text][Average BER vs. bit energy per user ($ E_b $) at 300\% loading]{
\includegraphics[width=0.46\textwidth]{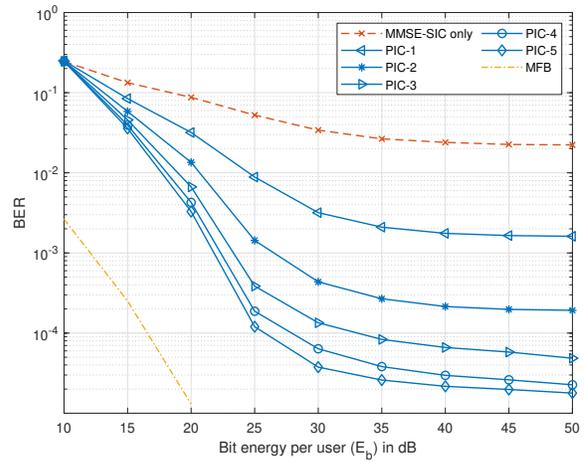}
\label{MUSA_iter_Eb}}
\caption{Average BER performance of PIC iterations for uplink MUSA with $ N_c = 16 $ and $ D = 32 $}
\label{MUSA_iter}
\end{figure}

PIC iterations for uplink MUSA improve the performance significantly. Fig. \ref{MUSA_iter_Nc} shows BER curves for different iterations with changing overloading at $ E_b = $ 40 dB with $ N_c = 16 $ and $ D = 32 $. It can be seen that PIC iterations, especially the first one, decreases BER of MMSE-SIC dramatically. Improvement in BER declines after a number of iterations. Similar characteristics are observed for same $ N_c $ and $ D $ for changing $ E_b $ when overloading is fixed to $ 300 \% $ in Fig. \ref{MUSA_iter_Eb}. It is seen that there is an error floor whose level gets considerably lower with PIC iterations. In addition, PIC iterations decrease the gap between MFB and MUSA especially at low $ E_b $ values.

\vspace{10mm}

\begin{figure}[h]
\centerline{\includegraphics[width=0.65\textwidth]{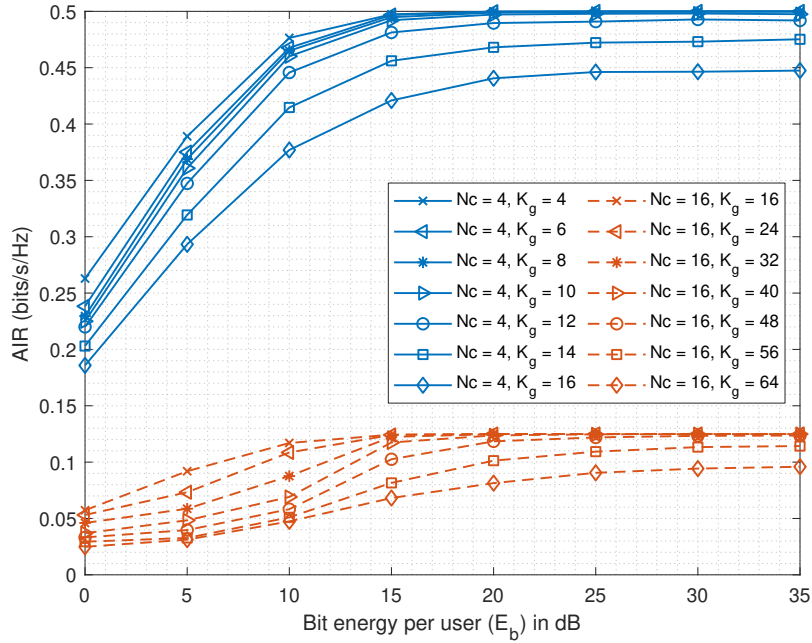}}
\vspace{-3mm}
\caption{Average AIR per user vs. bit energy per user ($ E_b $) for uplink MUSA with $ N_c = $ 4, 16 and $ D = $ 16}
\label{MUSA_AIR_Nc}
\vspace{5mm}
\end{figure}

AIR performance of MUSA was investigated at a fixed $ E_b $. In contrast, Fig. \ref{MUSA_AIR_Nc} shows average AIR per user curves with changing $ E_b $ where $ D $ is fixed to 16. AIR curves saturate at a point after some $ E_b $ value for both $ N_c = 4 $ and $ N_c = 16 $. Note that saturation point decreases with increasing overloading. It can be inferred that high overloading scenarios may cause loss in achievable rate for fixed $ N_c $ and $ D $ values.

\section{Conclusion}

In this paper, we have proposed a general framework on code-domain NOMA schemes for massive MIMO systems utilizing user-grouping based hybrid beamforming. Equivalent signal expressions for NOMA are obtained for both downlink and uplink transmission. Single-carrier is employed for uplink transmission as ISI is mitigated in beamspace. It is shown that classical receivers of two popular NOMA schemes, namely MUSA and SCMA, can be implemented without additional complexity. Furthermore, a novel receiver for uplink MUSA which uses PIC is proposed. AIR is introduced as an analysis tool in order to evaluate the performance of NOMA schemes with different NOMA parameters. Simulation results showed that NOMA schemes can work with very limited number of RF chains and they should be preferred if users are spatially close to each other. Flexible structure and low complexity of MUSA makes it preferable over SCMA, especially with the proposed PIC-aided uplink receiver whose effectiveness is clearly demonstrated. AIR and BER analysis of this receiver showed that high overloading can be realized with reasonable number of RF chains and code lengths.

\bibliographystyle{IEEEtran-mod}
\bibliography{References}

%




\end{document}